\documentclass[12pt]{article} %***

\usepackage[a4paper, total={6.5in, 8.5in}]{geometry}

\usepackage[sectionbib]{natbib}

\usepackage{amsmath}
\usepackage{amssymb}
\usepackage{stmaryrd}

\usepackage{tikz}
\usetikzlibrary{shapes,snakes}
\usetikzlibrary{positioning}

\usepackage{subcaption}
\usepackage{graphicx}

\usepackage[pdfencoding=auto,
            colorlinks = true,
            urlcolor  = blue,
            ]{hyperref}

\usepackage{multirow}

\usepackage{upgreek}

\usepackage{tablefootnote}

\usepackage{array}
\newcolumntype{C}[1]{>{\centering\let\newline\\\arraybackslash\hspace{0pt}}m{#1}}

\DeclareMathOperator*{\argmax}{arg\,max}

\DeclareMathOperator{\sigmoid}{\mathbf{sigmoid}}
\DeclareMathOperator{\gauss}{\mathcal{N}}
\DeclareMathOperator{\nat}{\mathbf{N}}
\DeclareMathOperator{\real}{\mathbf{R}}
\DeclareMathOperator{\bigOh}{\mathcal{O}}

\title{A State-Space Perspective on Modelling and \\Inference for Online Skill Rating}

\date{4th August 2023}
\author{Samuel Duffield\\ 
        \small{Normal Computing} \\ \\
        Samuel Power\footnote{\href{email:sam.power@bristol.ac.uk}{sam.power@bristol.ac.uk}}\\
        \small{School of Mathematics, University of Bristol, UK}\\ \\
        Lorenzo Rimella\\
        \small{Department of Mathematics and Statistics, Lancaster University, UK, and} \\
        \small{ESOMAS, University of Torino, IT}}

\begin{document}
 \maketitle

%%%%%%%%%%%%%%%%%%%%%%%%%%%%%%%%%%%%%%%%%%%%%%%%%%%%%%%%%%%%%%%%%%%%%%%%%%%%%%%%%%%%%%%%%%%%%%%%%%%%%%%%%%%%%%%%%%%%%%%%%%%%

\begin{abstract}
    We summarise popular methods used for skill rating in competitive sports, along with their inferential paradigms and introduce new approaches based on sequential Monte Carlo and discrete hidden Markov models. We advocate for a state-space model perspective, wherein players' skills are represented as time-varying, and match results serve as observed quantities. We explore the steps to construct the model and the three stages of inference: filtering, smoothing and parameter estimation. We examine the challenges of scaling up to numerous players and matches, highlighting the main approximations and reductions which facilitate statistical and computational efficiency. We additionally compare approaches in a realistic experimental pipeline that can be easily reproduced and extended with our open-source Python package, \href{https://github.com/SamDuffield/abile}{abile}.
\end{abstract}

\section{Introduction\label{sec:Introduction}}

In the quantitative analysis of competitive sports, a fundamental task is to estimate the skills of the different agents (‘players’) involved in a given competition based on the outcome of comparisons (‘matches’) between said players, often in an online setting. Skill rating is of paramount importance as it serves as a foundational tool for assessing and comparing the abilities of players and how they vary over time. By accurately quantifying skill levels, rating systems inform strategic decision-making and enhance the overall sporting level. Common applications include chess \citep{FIDE}, online gaming \citep{Herbrich2006}, education \citep{Pelanek2016}, tennis \citep{Kovalchik2016} and team-based sports like football \citep{Hvattum2010}, basketball \citep{Strumbelj2012} and many more \citep{Stefani2011,otting2020hot}. Skill ratings not only facilitate player ranking, but also serve as a basis for dynamic matchmaking and player seeding, ensuring that competitive matches are engaging and well-matched. Moreover, in professional sports, skill rating plays a pivotal role in talent scouting and gauging overall performance, providing data-driven insights for coaches, analysts, bettors and fanatics. As technology and statistical methodologies continue to advance, skill rating systems are expected to evolve further, benefiting an ever-widening spectrum of sports and competitive domains.

There are several established approaches to the task of skill estimation, including among others the Bradley-Terry model \citep{Bradley1952}, the Elo rating system \citep{Elo1978}, the Glicko rating system \citep{Glickman1999}, and TrueSkill \citep{Herbrich2006} each with various levels of complexity and statistical foundation. In these works, statistical models are often left implicit or not even present, hindering the incorporation of additional structure into estimation routines. Therefore, we propose a state-space modelling approach that separates model definition from inference and decomposes both aspects into their fundamental elements. This modularisation empowers practitioners to easily experiment with different model structures (e.g. binary vs non-binary match outcomes) and inferential paradigms (e.g. sequential Monte Carlo vs extended Kalman filter), ultimately selecting the most suitable combination for their specific needs.

Our key contributions can be summarised as follows:
\begin{itemize}
    \item We provide a unified state-space model framework for the skill rating, which encompasses and extends existing approaches by fostering the decoupling and modularisation of model design and general purpose inference. The state-space model represents a customisable blueprint, adaptable to specific needs (e.g. non-Gaussian dynamics, Bivariate-Poisson likelihoods, and more). We detail the relevant inference objectives (filtering, smoothing and parameter estimation) outlining their purpose, applications, and efficient implementation strategies.
    \item We rigorously formalise the \textit{factorial approximation} as the necessary bias introduced in order to scale computations up to realistic settings with numerous players while retaining statistical fidelity.
    \item We draw from the state-space model literature a suite of inferential methods. Including sequential Monte Carlo and finite-space methods that are completely novel in the setting of skill ratings and suffer from fewer sources of bias than previous model-based approaches, see Table~\ref{tab:comparison}.
    \item We walk through a realistic workflow of fitting and deploying the aforementioned skill rating tools on real datasets. We emphasise the key steps of model design and inference and explain how different components can be incorporated to improve prediction, robustness and interpretability.
    \item An open-source, lightweight and extensible Python package is provided to ensure complete reproducibility of the results and facilitate the use of the proposed and reviewed methodologies (\href{https://github.com/SamDuffield/abile}{github.com/SamDuffield/abile}).
\end{itemize}

The paper will proceed as follows. In Section \ref{sec:Model-Specification}, we describe a high-level formalism for state-space formulations of the skill estimation problem. In Section \ref{sec:Inference}, we outline the possible inference objectives for this problem and how they interact as well as the general structure of tractable inference, induced approximations and relevant complexity considerations. In Section \ref{sec:Concrete-Procedures}, we review a variety of concrete procedures for the skill estimation problem, combining probabilistic models and inference algorithms unifying existing approaches and introducing new ones. In Section \ref{sec:Experiments}, we perform some numerical experiments, demonstrating and contrasting some of the varied approaches to real data. Finally, in Section \ref{sec:Discussion}, we conclude with a discussion on general recommendations and extensions.

\section{Model Specification\label{sec:Model-Specification}}

Throughout, we will use various terminology which is specific to the skill estimation problem. We will use `sport' to denote a sport in the abstract (e.g. football), and will use `match' to denote a specific instance of this sport being played (e.g. a match between two football teams). All matches will be contested between two competing `players' (e.g. a football team is a `player'), one of whom is designated as the `home' player, and the other as the `away' player (even for sports in which there is no notion of `home ground' or `home advantage'). A `competition' refers to a specific sport, a collection of players of that sport, a set of matches between these players, and a set of results of these matches (e.g. the results of the English Premier League). Each match is also associated to a time $t \in \left[ 0, \infty \right)$ denoting when the match was played, which we call a `matchtime'. Driven by the popularity and simplicity of pairwise comparison \citep{gorgi2019analysis,wheatcroft2021forecasting}, we will focus on the aforementioned scenario where a `match' is played by only two `players', although the framework is general and easily extends to multiplayer comparisons which we later discuss in \ref{subsec:extensions}.
% We will provide a clear pipeline and a full open-source implementation allowing practitioners to review current methods while also integrating their own. %Nonetheless, we will highlight how our approach extends to multiplayer games, showcasing its versatility and generality.

Notationally, given an integer $N \in \nat$ we use $\left[ N \right]$ for the set of integers from $1$ to $N$. The total number of players in the competition is denoted $N \in \nat$, and $K \in \nat$ denotes the total number of matches played in the competition. We order the matchtimes as $t_{1} \leqslant t_{2} \leqslant \cdots \leqslant t_{K}$ (noting that matches may take place contemporaneously). Matches are then indexed in correspondence with this ordering, i.e. match $1$ took place at time $t = t_{1}$, etc.; we also adopt the convention that $t_{0}=0$. We explicitly model the skills of all players over the full-time window $[0, T]$, even if individual players may enter the competition at a later time.

We now discuss modelling choices. Our key interest is to infer the skills of players in a fixed competition, given access to the outcomes of matches which are played in that competition. We model player skills as taking values in a totally ordered set $\mathcal{X}$, i.e. skills are treated as ordinal, with the convention that higher skill ratings are indicative of more favourable match outcomes for a player.
The structure of the match outcomes (denoted $y_k$ for the outcome of the $k$th match) depends on the application and model design. Most simply and commonly, they can be represented by a finite, discrete set $y_k \in \mathcal{Y}$, e.g. $\mathcal{Y} = \left\{ \text{draw}, \text{home win}, \text{away win}\right\}$. Other structures are also readily accommodated, e.g. recording the number of points scored by each team as $y_k \in \mathcal{Y} = \nat^2$.

Our generic model will consist of the players' skills and the match outcomes; we detail here the construction of the full joint likelihood. Players' skills are allowed to vary with time, and we write $x_{t}^{i} \in \mathcal{X}$ for the skill value of the $i$th player at time $t$ and in an abuse of notation, we will also use $x_k^{i}$ to denote the skill value of the $i$th player at observation time $t_k$. We will, where possible, index skills with letters $t$ or $k$ so that the continuous/discrete nature of the time index is clear from the context. For discrete-time indices, we also make use of the notation $x_{0:k}$ or $y_{1:k}$ to denote the joint vector of skills or observations at discrete times.

We assume that the initial skills of the players are all drawn mutually independently of one another, i.e. for $i \in \left[ N \right]$, $x_{0}^{i} \sim m_{0}^{i}$ independently for some initial distribution $m_0^i$. We also assume that the evolution of each player's skill over time is independent of one another. We further assume that each of these evolutions is Markovian, i.e. for each $i \in \left[ N \right]$, there is a semigroup of $\mathcal{X}$-valued Markov transition kernels $\left( M_{t,t^\prime}^{i} : t \leqslant t^\prime \right)$ such that $x_{t^\prime}^{i} \mid x_t^{i} \sim M_{t, t^\prime}^{i} \left(x_t^{i}, \cdot \right)$
where $t \leqslant t^\prime$ represent matchtimes. Again, we will often abuse notation to write $M_{k,k^\prime}^{i} = M_{t_{k},t_{k^\prime}}^{i}$ with $t_k \leqslant t_{k^\prime}$ and the nature of the index clear from context.

\begin{figure}[httb!]
\centering
\begin{subfigure}{0.45\textwidth}
\resizebox{0.9\textwidth}{!}{
\begin{tikzpicture}[-latex, auto, node distance =3 cm and 3cm ,on grid ,
    state/.style ={ circle ,top color =white , draw , text=blue,font=\bfseries\Large , 
    minimum width =2cm}]

\node[state] (Xt-1) at (0,0) {$x_{{k-1}}$};
\node[state] (Xt) [right =of Xt-1] {$x_{k}$};
\node[state] (Xt+1) [right =of Xt  ] {$x_{{k+1}}$};

\node[draw=none] (none)[above =of Xt-1] {};
\node[state] (Yt-1)[below =of Xt-1] {$y_{k-1}$};
\node[state] (Yt)[below =of Xt] {$y_{k}$};
\node[state] (Yt+1)[below =of Xt+1] {$y_{k+1}$};

\node[draw=none] (fYt-1)[below =of Yt-1] {};
\node[draw=none] (fYt)  [below =of Yt  ] {};
\node[draw=none] (fYt+1)[below =of Yt+1] {};

\node[draw=none] (ffYt-1)[below =of fYt-1] {};
\node[draw=none] (ffYt)  [below =of fYt  ] {};
\node[draw=none] (ffYt+1)[below =of fYt+1] {};

{};

\path (-2,0) edge [bend left = 0] node {} (Xt-1);
\path (Xt-1) edge [bend left = 0] node {} (Xt);
\path (Xt) edge [bend left = 0] node {} (Xt+1);
\path (Xt+1) edge [bend left = 0] node {} (9,0);

\path (Xt-1) edge [bend left = 0] node {} (Yt-1);
\path (Xt) edge [bend left = 0] node {} (Yt);
\path (Xt+1) edge [bend left = 0] node {} (Yt+1);

\end{tikzpicture}
}
\end{subfigure}
\begin{subfigure}{0.45\textwidth}
\hfill
\resizebox{0.9\textwidth}{!}{
\begin{tikzpicture}[-latex, auto, node distance =3 cm and 3cm ,on grid ,
    state/.style ={ circle ,top color =white , draw , 
    text=blue,font=\bfseries\large, minimum width =2 cm}]

\node[state] (Xt-11) at (0,0) {$x_{{k-1}}^1$};
\node[state] (Xt1) [right =of Xt-11]   {$x_{k}^1$};
\node[state] (Xt+11) [right =of Xt1  ] {$x_{{k+1}}^1$};

\node[state] (Yt-1)[below =of Xt-11]   {$y_{k-1}$};
\node[state] (Yt)[below =of Xt1]       {$y_{k}$};
\node[state] (Yt+1)[below =of Xt+11]   {$y_{k+1}$};

\node[state] (Xt-12) [above =of Xt-11] {$x_{{k-1}}^2$};
\node[state] (Xt2) [above =of Xt1]     {$x_{k}^2$};
\node[state] (Xt+12) [above =of Xt+11] {$x_{{k+1}}^2$};

\node[state] (Xt-13) [above =of Xt-12] {$x_{{k-1}}^3$};
\node[state] (Xt3) [above =of Xt2]     {$x_{k}^3$};
\node[state] (Xt+13) [above =of Xt+12] {$x_{{k+1}}^3$};

\node[state] (Xt-14) [above =of Xt-13] {$x_{{k-1}}^4$};
\node[state] (Xt4) [above =of Xt3]     {$x_{k}^4$};
\node[state] (Xt+14) [above =of Xt+13] {$x_{{k+1}}^4$};
\node[state] (Xt+14) [above =of Xt+13] {$x_{{k+1}}^4$};

\path (-2, 0) edge node {} (Xt-11);
\path (-2, 3) edge node {} (Xt-12);
\path (-2, 6) edge node {} (Xt-13);
\path (-2, 9) edge node {} (Xt-14);

\path (Xt-11) edge node {} (Xt1);
\path (Xt-12) edge node {} (Xt2);
\path (Xt-13) edge node {} (Xt3);
\path (Xt-14) edge node {} (Xt4);

\path (Xt1) edge node {} (Xt+11);
\path (Xt2) edge node {} (Xt+12);
\path (Xt3) edge node {} (Xt+13);
\path (Xt4) edge node {} (Xt+14);

\path (Xt+11) edge node {} (9, 0);
\path (Xt+12) edge node {} (9, 3);
\path (Xt+13) edge node {} (9, 6);
\path (Xt+14) edge node {} (9, 9);

\path (Xt-11) edge [bend left = 40] node {} (Yt-1);
\path (Xt-12) edge [bend left = 40] node {} (Yt-1);

\path (Xt1) edge [bend left = 40] node {} (Yt);
\path (Xt3) edge [bend left = 40] node {} (Yt);

\path (Xt+12) edge [bend left = 40] node {} (Yt+1);
\path (Xt+14) edge [bend left = 40] node {} (Yt+1);

\end{tikzpicture}
}
\end{subfigure}
\caption{Left: conditional independence structure of an SSM. Right: conditional independence structure of an fSSM \eqref{eq:joint-lik}, with $N=4$ and pairwise observation model.}\label{fig:fSSM}
\end{figure}
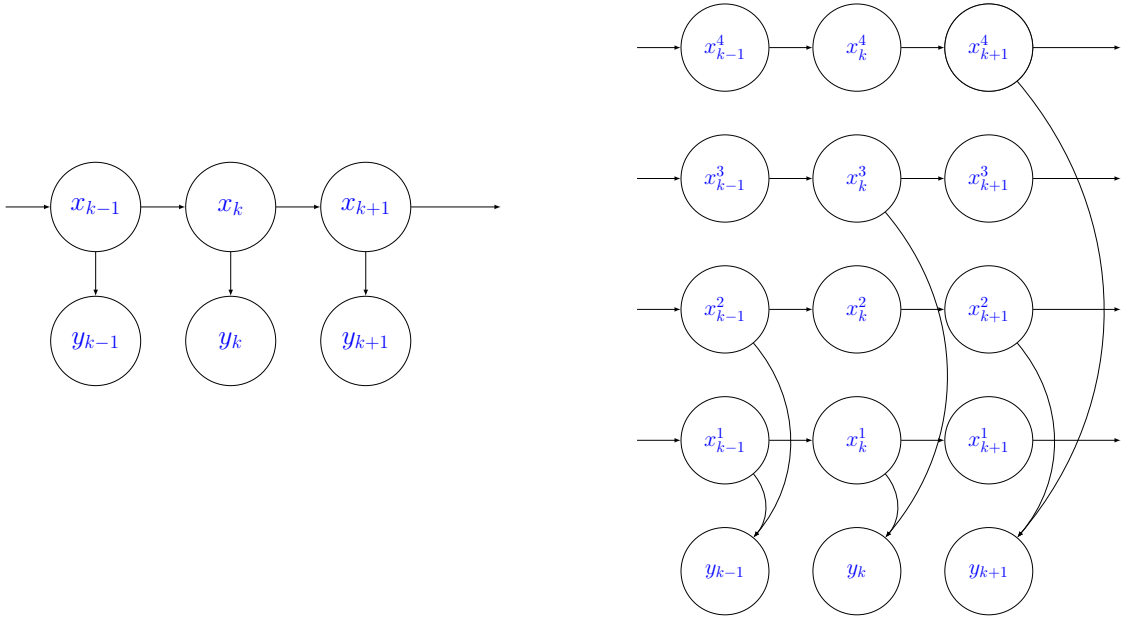

For the $k$th match, we write $h(k), a(k) \in \left[ N \right]$ for the indices of the home and away players in that match, and write $y_{k}\in\mathcal{Y}$ for the outcome of the match. We assume that given the skills of players $h(k)$ and $a(k)$ at time $t_{k}$, the match's outcome $y_{k}$ is conditionally independent of all other player skills and match outcomes (depicted in Fig.~\ref{fig:fSSM}), and is drawn according to some probability distribution $G_{k} \left( y_{k} \mid x_{{k}}^{h \left( k \right)}, x_{{k}}^{a \left( k \right)}\right)$. We again emphasise that practitioners are completely free to choose the observation model that suits their application, whether sigmoidal, Poisson, or otherwise. The only structural condition imposed upon $G_k$ is that the observation depends only on the few players that are involved in the match, i.e. $\{h(k),a(k)\}$. 

% or $C(k)$ for the multiplayer one. Note that the observation $y_k$ depends only on the current skill levels of players $\{h(k),a(k)\}$, but we can easily extend this formulation to a set of players $C(k)$ resulting in $G_{k} \left( y_{k} \mid x_{{k}}^{C(k)}\right)$.  

Assembling these various assumptions, it follows that the joint law of all players' skills on all matchtimes, and of all match results, is thus given by the following

\begin{equation}\label{eq:joint-lik} 
    \begin{split}
        \mathbf{P} \left( x_{0:K}^{\left[ N \right]}, y_{1:K} \right) = &\prod_{i \in \left[ N \right]} \left\{ m_{0}^{i} \left( x_{0}^{i} \right) \cdot \prod_{k=1}^K M_{{k-1}, {k}}^{i} \left(x_{{k-1}}^{i}, x_{{k}}^{i} \right) \right\} \\
        &\cdot \prod_{k=1}^K G_{k} \left(y_{k} \mid x_{{k}}^{h \left( k \right)},x_{{k}}^{a \left( k \right)}\right). 
    \end{split}
\end{equation}

% \begin{equation}\label{eq:joint-lik} 
% \begin{split}
%     \mathbf{P} \left( x_{0:K}^{\left[ N \right]}, y_{1:K} \right) = \prod_{i \in \left[ N \right]} & \left\{ m_{0}^{i} \left( x_{0}^{i} \right) \cdot \prod_{k=1}^K M_{{k-1}, {k}}^{i} \left(x_{{k-1}}^{i}, x_{{k}}^{i} \right) \right\} 
%     \cdot \prod_{k=1}^K G_{k} \left(y_{k} \mid x_{{k}}^{h \left( k \right)},x_{{k}}^{a \left( k \right)}\right). 
% \end{split}
% \end{equation}

Some simplifications which we will make in all subsequent examples are that i) we will model the initial laws of all players' skill as being identical across players (i.e. $m_0^{i}$ will not depend on $i$), ii) the dynamics of all players' skills will also be identical across players, i.e. for $t < t^\prime$, we can write $M_{t, t^\prime}^{i} \left( x, x^\prime \right) = M_{t, t^\prime} \left( x, x^\prime \right)$, and iii) the observation model $G_k$ will not depend on $k$ (although we will still use the notation $G_k$ to emphasise dependence on $y_k$). These simplifications are made for ease of presentation, and deviations from each of these simplifications are typically straightforward to accommodate in the algorithms which we present. Such deviations are often relevant in practice, e.g. different match types (e.g. 3-set vs 5-set tennis).

There are also a number of model features on which we insist, namely i) we insist on modelling player skills as evolving in continuous time, and ii) we insist on the possibility of observations which occur at irregularly-spaced intervals in time. We do this because these settings are of practical relevance, and because they pose particular computational challenges which deserve proper attention.

Terminologically, we use the term `state-space model' (SSM) to denote a model, such as \eqref{eq:joint-lik}, in which there is an unobserved state $x$, taking values in a general space, which evolves in time according to a Markovian evolution, and is observed indirectly. We use the term `hidden Markov model' (HMM) to refer to an SSM where the unobserved state $x$ takes values on a finite state space. The term factorial state-space model (fSSM, or indeed fHMM) refers to a state-space model in which the state $x$ is naturally partitioned into a collection of sub-states, each of which evolves independently of one another \citep{ghahramani1995factorial}, see Fig.~\ref{fig:fSSM}.

\section{Inference\label{sec:Inference}}

In this section, we discuss the problem of inference in the skill rating problem, which features of the problem one might seek to understand, and how one might go about representing these features.

\subsection{Inference Objectives}

Broadly speaking, inference in general state-space models tends to involve the solution of three related tasks, presented in (roughly) increasing order of complexity:
\begin{enumerate}
    \item Filtering: inferring the current latent states, given the observations
    thus far, as well as prediction
    \begin{align*}
        \mathrm{Filter}_k \left(x_k\right) := \mathbf{P} \left(x_{k} \mid y_{1:k}  \right),
        \quad \mathrm{Predict}_{k+1 \mid k} \left(x_{k+1} \right) = \mathbf{P} \left(x_{k+1} \mid y_{1:k} \right).
    \end{align*}
    for $t_{k}<t_{k+1}$.
    \item Smoothing: inferring past latent skills, given
    the observations thus far
    \begin{align*}
        \mathrm{Smooth}_{0:K|K} \left( x_{0:K} \right) = \mathbf{P} \left( x_{0:K}\mid {y}_{1:K} \right).
    \end{align*}
    with $\mathrm{Smooth}_{k,k+1|K} \left( x_{k}, x_{k+1} \right) = \mathbf{P} \left( x_{k}, x_{k+1} \mid {y}_{1:K} \right)$ and
    $\mathrm{Smooth}_{k|K} \left( x_{k} \right) = \mathbf{P} \left( x_{k}\mid {y}_{1:K} \right)$ .
    \item Parameter Estimation: when the dynamical  and/or observational structure of the model depends on unknown parameters $\theta$, one can calibrate these models based on the observed data by e.g. maximum likelihood estimation:
    \begin{align*}
        \argmax_{\theta \in \Theta } \mathbf{P} \left( {y}_{1:K} \mid \theta\right) ,
    \end{align*}
    where $\mathbf{P}( {y}_{1:K} \mid \theta) = \int \mathbf{P} \left( x_{{0:K}}^{\left[N\right]}, {y}_{1:K} \mid \theta \right) \, \mathrm{d} x_{0:K}^{\left[ N \right]}$. Note that for filtering and smoothing, $\theta$ is treated as constant, hence its omission from the preceding descriptions.
\end{enumerate}

Depending on the application in question, each of these tasks can be of more or less interest. In our setting, because we are interested in using our model to make real-time decisions, filtering is directly relevant towards informing those decisions. However, this does not mean that we can immediately ignore the other two tasks. Firstly, without an accurate estimate of the parameters $\theta$ which govern the dynamical and observation models for our process of interest, our estimate of the filtering distribution can be badly misspecified. As a result, without incorporating some elements of parameter estimation, inference of the latent states can be quite poor. Moreover, obtaining good estimates of these model parameters tends to require developing an understanding of the full trajectory of the latent process; indeed, many algorithms for parameter estimation in SSMs require some form of access to the smoothing distribution, which in turn requires the filtering distributions. As such, the three tasks are deeply interconnected.

% In Section \ref{sec:Concrete-Procedures}, we will explicitly consider computational methods for addressing these inference objectives. For the problems considered in this paper, it is rare that the true filtering or smoothing distributions can even be represented, and as such, approximations will be adopted. It bears mentioning that the `richness' of the chosen approximation will play an important role in determining how well the original inference goals are achieved.
% This will be treated further in Section \ref{sec:Concrete-Procedures}.

\subsection{Techniques for Filtering} \label{sec:filtering}

We first present some generalities on algorithmic approaches to the filtering and smoothing problem, to contextualise the forthcoming developments. For a general SSM on $\mathcal{X}$ with transitions $M_{t,t^\prime}$ and observations $G_k$, the following abstract filtering recursions hold
\begin{align*}
    \mathrm{Predict}_{k+1|k} \left( x_{k+1}\right) &= \int \mathrm{Filter}_{k} \left(x_{k} \right) \cdot M_{k,k+1} \left(x_{k}, x_{k+1} \right) \, \mathrm{d}x_{k}, \\
    \mathrm{Filter}_{k+1} \left(x_{k+1}\right) &\propto {\mathrm{Predict}_{k+1|k} \left( x_{k+1}\right) \cdot G_{k+1} \left(x_{k+1} \right)},
\end{align*}
or more suggestively,
\begin{align*}
    \text{} \mathrm{Predict}_{k+1|k} & = \mathsf{Propagate} \left(\mathrm{Filter}_k; M_{k,k+1}\right),\\
    \mathrm{Filter}_{k+1} &= \mathsf{Assimilate} \left(\mathrm{Predict}_{k+1|k}; G_{k+1}\right),
\end{align*}
where we note that the operators $\mathsf{Propagate}$ and $\mathsf{Assimilate}$ act on probability measures. Note that the $\mathsf{Propagate}$ operator can also readily be applied to times which are not associated with matches; this can be useful for forecasting purposes.

The same recursions also allow for computation of the likelihood of all observations so far, using that $\mathbf{P} \left( {y}_{1:k+1} \right) = \mathbf{P} \left( {y}_{1:k} \right) \cdot \mathbf{P} \left( y_{k+1} \mid {y}_{1:k} \right)$, the last term carries the interpretation of a predictive likelihood or in the skill-rating setting, match outcome predictions.

% Various algorithms for approximate filtering have been derived by approximating each of these updates in turn. The computational cost varies depending on the algorithm being used; we will discuss these on a case-by-case basis. 

\subsection{Techniques for Smoothing}\label{sec:smoothing}

Similarly to filtering, there are `backward' recursions which characterise the smoothing laws:  if no observations occur in the interval $\left( t_k, t_{k+1} \right)$, then
\begin{align*}
    \mathrm{Smooth}_{k, k + 1 | K} \left( x_{k}, x_{k+1} \right) & = \frac{\mathrm{Filter}_{k} \left( x_{k} \right) \cdot M_{k, k + 1} \left(x_{k}, x_{k + 1} \right) \cdot \mathrm{Smooth}_{k + 1 | K} \left( x_{k + 1}\right)}{\mathrm{Predict}_{k+1|k}(x_{k+1})}, \\
    \mathrm{Smooth}_{k|K} \left(x_{k} \right) & =\int \mathrm{Smooth}_{k,k + 1|K} \left( x_{k}, x_{k+1} \right) \, \mathrm{d}x_{k+1},
\end{align*}
or
\begin{align*}
    \mathrm{Smooth}_{k,k+1|K} &= \mathsf{Bridge} \left(\mathrm{Filter}_k, \mathrm{Smooth}_{k+1|K}; M_{k,k+1}\right)\\
    \mathrm{Smooth}_{k|K} & =\mathsf{Marginalise} \left( \mathrm{Smooth}_{k,k+1|K}; k\right).
\end{align*}
%
% where, as for filtering, the operators act on probability measures, and noting that $M_{k, k+1}$ is interpreted as a density, rather than as a Markov kernel.
% Likewise, several algorithms for approximate smoothing are built upon approximation of these recursions.

Smoothing between observation times is possible using the same approach, i.e. for $t_k \leqslant t_{k^\prime} \leqslant t_{k+1}$ where $t_{k^\prime}$ is not associated with an observation, we have
\begin{align*}
    \mathrm{Smooth}_{k^\prime,k+1|K} = \mathsf{Bridge} \left(\mathrm{Predict}_{k^\prime | k}, \mathrm{Smooth}_{k+1|K}; M_{k^\prime,k+1}\right).
\end{align*}

In general, exact implementation of any of $\left\{ \mathsf{Propagate}, \mathsf{Assimilate}, \mathsf{Bridge}, \mathsf{Marginalise}\right\}$ tends to only be possible in models with substantial conjugacy properties, due to the general difficulty of integration and representation of probability measures of even moderate complexity. If one insists on an exact implementation of all of these operations, then one tends to be restricted to working with linear Gaussian SSMs or HMMs of moderate size. As such, practical algorithms must often make approximations which restore some level of tractability to the model.

Observe also that, given the filtering distributions, the smoothing recursions require no further calls to the likelihood term $G_k$, which can be noteworthy in the case that the likelihood is computationally expensive or otherwise complex.

\subsection{Standing Approximations and Reductions}

We are interested in developing procedures for filtering, smoothing, and parameter estimation involving i) many players, i.e. $N \to \infty$, and ii) many matches, i.e. $K \to \infty$. We will therefore focus on procedures for which the computational cost scales at most \textit{linearly} in each of $N$ and $K$.

%We are interested in developing procedures for filtering, smoothing, and parameter estimation whose complexity scales well with respect to the parameters of interest for skill rating. In particular, we want to be able to process competitions involving i) many players, i.e. $N \to \infty$, and ii) many matches, i.e. $K \to \infty$. We will therefore focus on procedures for which the computational cost scales at most \textit{linearly} in each of $N$ and $K$.

\subsubsection{Decoupling Approximation}\label{sec:decoupling_appprox}

In seeking procedures with stable behaviour as the number of players grows, there is one approximation which seems to be near-universal in the setting of pairwise comparisons, and beyond, namely that the filtering (and smoothing) distributions over all of the players' skills are well-approximated by a decoupled representation. Using superscripts to index player-specific distributions (e.g. $\mathrm{Filter}_k^i$ denoting the filtering distribution for the skill of player $i$ at time $t_k$, and so on), this corresponds to the approximation
\begin{align*}
    \mathrm{Filter}_k = \mathbf{P} \left(x_{k}^{\left[N\right]} \mid y_{1:k} \right) & \approx \prod_{i\in\left[N\right]} \mathbf{P} \left(x_{k}^{i}\mid y_{1:k} \right) = \prod_{i\in\left[N\right]} \mathrm{Filter}_k^i  \\
    \mathrm{Smooth}_{\circ \mid K} = \mathbf{P} \left( x_{\circ}^{\left[N\right]} \mid y_{1:K}  \right) & \approx \prod_{i \in \left[ N \right]} \mathbf{P} \left(x_{\circ}^{i} \mid y_{1:K} \right) = \prod_{i \in \left[ N \right]} \mathrm{Smooth}_{\circ \mid K}^i,
\end{align*}
where we have $\circ = \left\{ k \right\}, \left\{ k, k+1 \right\}$ or $\{ 0: K\}$ for the various marginal and joint smoothing objectives. This approximation is motivated by the practical difficulty of representing large systems of correlated variables and is further supported by the standing assumption that players' skills evolve independently of one another a priori. 

% The quality of such approximations depends heavily on the ability to control the strength of interactions between players, that is, the sensitivity of the conditional law of any one player's skill to perturbations in any single other player's skill. If such control is possible (which one expects for large-scale, high-frequency competitions, with weakly-informative match outcomes), then one can rigorously establish that the decoupling approximation has good fidelity to the true filtering law \citep{rebeschini2015can, rimella2022exploiting}.
% Whether richer approximations of the filtering and smoothing laws are practically feasible and worthwhile remains to be seen. We thus focus hereafter on inferential paradigms which adopt this decoupling approximation.

\subsubsection{Match Sparsity}\label{sec:match_sparsity}

Our general formulation of the joint model includes more information than is strictly necessary. Due to the conditional independence structure of the model, one sees that instead of monitoring the skills of all players during all matches, it is sufficient to keep track of only the skills of players at times when they are playing in matches. Reformulating a joint likelihood which reflects this simplicity requires the introduction of some additional notation, but dramatically reduces the cost of working with the model, and is computationally crucial.

To this end, for $i \in \left[ N \right]$, write $L^{i} \subseteq \{0, \dots, K\}$ for the ordered indices of matches in which player $i$ has played, and for $\ell^i \in L^i$, write $\ell^{i,-}$ for the element of $L^i$ immediately before $\ell^i$, i.e. $\ell^{i,-} = \sup \{ \ell \in L^{i} : \ell < \ell^i \}$. It then holds that
\begin{align}\label{eq:joint_lik_sparse}
    \mathbf{P} \left(\left\{ x_{\ell^i}^i : \ell^i \in L^i, i \in {[N]} \right\} , {y}_{1:K} \right) & = \prod_{i \in \left[ N \right]} \left\{ m_{0} \left( {x}_{0}^{i} \right) \cdot \prod_{\ell^i \in  L^{i}} M_{\ell^{i,-}, \ell^i} \left( {x}_{\ell^{i,-}}^{i}, {x}_{\ell^i}^{i} \right) \right\}
    \nonumber
    \\
    & \qquad \cdot \prod_{k=1}^K G_k \left(y_{k} \mid x_{{k}}^{h \left(k\right)}, x_{{k}}^{a \left(k\right)}\right). 
\end{align}
Whilst our original likelihood \eqref{eq:joint-lik} contained $\bigOh (N\cdot K)$ terms, this new representation involves only $\bigOh ( N + K)$ terms. Given that the competition consists of $N$ players and $K$ matches, we see that this representation is essentially minimal. %Interestingly, the switch to a multiplayer scenario is again straightforward and simply requires accounting for multiple players in the observation model as the sets $L^i$'s will have the same form.

\subsubsection{Pairwise Updates} \label{sec:pairwise_updates}

A consequence of these two features is that when carrying out filtering and smoothing computations, only sparse access to the skills of players is required. In particular, consider assimilating the result of the $k$th match into our beliefs of the players' skills. Since this match involves the players ${h(k)}$ and ${a(k)}$, which we refer to as $h$ and $a$, we have $k \in L^h \cap L^a$ and the filtering update requires only the following steps:
\begin{enumerate}
    \item Compute the last matchtime indices on which the two players played $k^{h,-} \in L^h$ and $k^{a,-} \in L^a$, and retrieve the filtering distributions of the two players' skills on these matchtimes, i.e. $\mathrm{Filter}_{k^{h,-}}^{h}$ and $\mathrm{Filter}_{k^{a,-}}^{a}$ respectively.
    \item Compute the predictive distributions of the two players' skills prior to the current matchtime by propagating them through the dynamics, i.e.  $M_{k^{h,-}, k}$ for the home player and $M_{k^{a,-},k}$ for the away player, compute
    \begin{align*}
        \left(
            \begin{array}{c}
                \mathrm{Predict}_{k \mid k^{h,-}}^{h}\\
                \mathrm{Predict}_{k \mid k^{a,-}}^{a}
            \end{array}
        \right) &= \left(
            \begin{array}{c}
                \mathsf{Propagate}\left(\mathrm{Filter}_{k^{h,-}}^{h}; M_{k^{h,-}, k}\right) \\
                \mathsf{Propagate}\left(\mathrm{Filter}_{k^{a,-}}^{a}; M_{k^{a,-},k}\right)
            \end{array}
        \right).
    \end{align*}
    \item Compute the current filtering distributions of the two players' skills immediately following the current matchtime by assimilating the new result
    \begin{align*}
        \mathrm{Filter}_{k}^{h,a}&=\mathsf{Assimilate}\left(\left(
            \begin{array}{c}
                \mathrm{Predict}_{k \mid k^{h,-}}^{h} \\
                \mathrm{Predict}_{k \mid k^{a,-}}^{a}
            \end{array}
        \right); G_{k}\right),
    \end{align*}
    where $\mathsf{Assimilate}$ denotes a Bayesian procedure that converts predictive distributions for a pair of players and a likelihood into a joint filtering distribution.
    \item Marginalise to regain the factorial approximation
    \begin{align*}
        \left(
            \begin{array}{c}
                \mathrm{Filter}_{k}^{h} \\
                \mathrm{Filter}_{k}^{a}
            \end{array}
        \right) &= \left(
            \begin{array}{c}
                \mathsf{Marginalise}\left (\mathrm{Filter}_{k}^{h,a}; h \right ) \\
                \mathsf{Marginalise}\left (\mathrm{Filter}_{k}^{h,a}; a \right )
            \end{array}
        \right),
    \end{align*}
    where $\mathsf{Marginalise}$ is used for a marginalisation in space and not in time.
\end{enumerate}
Note briefly that if at a specific time, multiple matches are being played between a disjoint set of players, then this structure implies that the results of all of these matches can be assimilated independently and in parallel; for high-frequency competitions in which many matches are played simultaneously, this is an important simplification. A key takeaway from this observation is then that the cost of assimilating the result of a single match is independent of both $N$ and $K$. 

Similar benefits are also available for smoothing updates. Indeed, the benefits of sparsity are even more dramatic in this case, since the smoothing recursions (Section~\ref{sec:smoothing}) decouple \textit{entirely} across players, implying that all smoothing distributions can be computed independently and in parallel. Given access to sufficient compute parallelism, this implies the possibility of computing these smoothing laws in time $\bigOh \left(\max \left\{ \mathbf{card} (L^i) : i \in  \left[ N \right] \right\} \right)$, which is potentially much smaller than the serial complexity of $\bigOh \left( K \right)$. For example, if each player is involved in the same number of matches ($\sim K / N$), then the real-time complexity is reduced by a factor of $N$.

%Again the multiplayer scenario is not too complex. Indeed, we can just follow steps 1 to 5 in the same way while being careful on considering all players in $C(k)$. This will require: retrieving the latest match times of the players and the corresponding  filtering distributions, compute predictive distributions for the players, assimilate the new result and then marginalising back to a factorial approximation. In particular, the $\mathsf{Assimilate}$ procedure in step 4 requires mapping to a joint distribution on all players in $C(k)$, i.e. producing $\mathrm{Filter}_{k}^{C(k)}$.    

For ease of exposition, in what follows, we will present all considered methods with notation corresponding to the $\bigOh (N \cdot K)$ posterior \eqref{eq:joint-lik}, rather than the memory-efficient $\bigOh (N + K)$ posterior \eqref{eq:joint_lik_sparse} which should be applied in practice, e.g. we will describe the $\mathsf{Propagate}$ step as taking us from time $k-1$ to time $k$, and so on.

\subsection{Techniques for Parameter Estimation}\label{subsec:ParamEst}

Recall the goal of parameter estimation is to infer the unknown static (i.e. non-time-varying) parameters $\theta$ of the state-space model. A complete discussion of parameter estimation in state-space models would be time-consuming. In this work, we simply focus on estimation following the principle of maximum likelihood, due to its generality and compatibility with typical approximation schemes. Careful discussion of other approaches (e.g. composite likelihoods, Bayesian estimation, etc.; see e.g. \cite{varin2008pairwise, varin2011overview, andrieu2010particle}) is omitted for reasons of space. Additionally, we limit our attention to offline parameter estimation, where we have access to historical match outcomes and look to find static parameters $\theta$ which model them well, with the goal of subsequently using these parameters in the online setting (i.e. filtering). Techniques for online parameter estimation (with a focus on particle methods) are reviewed in \cite{kantas2015particle}.

% Given the general intractability of the filtering and smoothing distributions, it should not be particularly surprising that the likelihood function of a state-space model is also typically unavailable. Fortunately, many approximation schemes for filtering and smoothing also enable the computation of an approximate or surrogate likelihood, which can then be used towards parameter estimation. That is, in addition to approximating the filtering distribution (for example), an algorithm may also provide access to a tractable approximation $\hat{\mathbf{P}} \left(y_{1:K} \mid \theta \right)$, which can then be optimised directly by numerical methods.

A popular and general option for parameter estimation is an expectation-maximisation (EM) strategy for maximising the likelihood $\mathbf{P} \left(y_{1:K} \mid \theta \right)$; see e.g. \cite{neal1998view} or Chapter 14 in \cite{Chopin2020} for overview. EM is an iterative approach, with each iteration consisting of two steps, known as the E-step and M-step respectively. The usual expectation step (`E-step') consists of taking the current parameter estimate $\hat{\theta}$, using it to form the smoothing distribution $\mathbf{P} \left(x_{0:K}^{\left[N\right]} \mid y_{1:K}, \hat{\theta}\right)$, and constructing the surrogate objective function
\begin{align}\label{eq:em_q}
    \mathbf{Q} \left( \theta \mid \hat{\theta} \right) := \int \mathbf{P} \left(x_{0:K}^{\left[ N \right]} \mid y_{1:K},\hat{\theta} \right) \cdot \log \mathbf{P}\left( x_{0:K}^{\left[N\right]},  y_{1:K} \mid \theta \right) \, \mathrm{d}x_{0:K}^{\left[N\right]},
\end{align}
which is a lower bound on $\log\mathbf{P} \left(y_{1:K} \mid \theta \right)$. The maximization step (`M-step') then consists of deriving a new estimate by maximising this surrogate objective. When these steps are carried out exactly, each iteration of the EM algorithm is then guaranteed to ascend the likelihood $\mathbf{P}(y_{1:K} \mid \theta)$, and will thus typically yield a local maximiser when iterated until convergence.

Depending on the complexity of the model at hand, it can be the case that either the E-step, the M-step, or both cannot be carried out exactly. For the models considered in this work, the intractability of the smoothing distribution means that the E-step cannot be carried out exactly. As such, we will simply approximate the E-step by treating our approximate smoothing distribution as exact (noting that this compromises EM's usual guarantee of ascending the likelihood function). Similarly, when the M-step cannot be carried out in closed-form, one can often approximate the maximiser of $\theta \mapsto \mathbf{Q} \left( \theta \mid \hat{\theta} \right)$ through the use of numerical optimisation schemes. 
%For a broad perspective on the EM algorithm and its approximations, we recommend \cite{neal1998view}.

In some cases, constructing the smoothing distribution in a way that provides a tractable M-step may be expensive, and it is cheaper to directly form the log-likelihood gradient. By Fisher's identity (see e.g. \cite{del2010forward}), this gradient takes the following form 
\begin{align}\label{eq:em_grad}
    \nabla_{\theta} \log\mathbf{P} \left(y_{1:K} \mid \theta \right) = \int \mathbf{P} \left(x_{0:K}^{\left[N\right]} \mid y_{1:K},\theta \right) \cdot \nabla_\theta \log \mathbf{P} \left(x_{0:K}^{\left[N\right]}, y_{1:K} \mid \theta \right) \, \mathrm{d}x_{0:K}^{\left[N\right]}.
\end{align}
As such, when the smoothing distribution is directly available, this offers a route to implementing a gradient method for optimising $\log \mathbf{P} \left(y_{1:K} \mid \theta \right)$. When only approximate smoothing distributions are available, one obtains an inexact gradient method, which may nevertheless be practically useful.

It is important to note that the logarithmic nature of the expectations in (\ref{eq:em_q}-\ref{eq:em_grad}) means that for many models, the components of $\theta$ can be treated independently. For example, when the parameters which influence $m_0$, $M_{t,t^{\prime}}$  and $G_k$  are disjoint, the intermediate objective $\mathbf{Q} \left( \theta \mid \hat{\theta} \right)$ will be separable with respect to this structure, which can simplify implementations. Moreover, depending on the tractability of solving each sub-problem, one can seamlessly blend the analytic maximisation of some parameters with gradient steps for others, as appropriate.

\begin{table}[httb]
\caption{\label{tab:comparison}Considered approaches and their features. All approaches are linear in the number of players $\bigOh(N)$ and the number of matches $\bigOh(K)$.
}
% }
\centering
% \addtolength{\tabcolsep}{-3pt}
% {
% \resizebox{1.2\textwidth}{0.075\textheight}{
\resizebox{\textwidth}{!}{
\begin{tabular}{|l|c|c|c|c|c|}
\hline 
Method & Skills & Filtering & Smoothing &
{\begin{tabular}[c]{@{}c@{}}Parameter\\ Estimation\end{tabular}}
&
{\begin{tabular}[c]{@{}c@{}}Sources of Error\\ (Beyond Factorial)\end{tabular}}
\tabularnewline
\hline 
\hline 
% Bradley-Terry & Static, General & N/A & N/A & MLE & N/A\tabularnewline
% \hline 
Elo & Continuous & Location , $\bigOh (1)$  & N/A & N/A & Not model-based\tabularnewline
\hline 
Glicko & Continuous & Location and Spread, $\bigOh (1)$  & Location and Spread, $\bigOh (1)$  & N/A & Not model-based\tabularnewline
\hline
Extended Kalman & Continuous & Location and Spread, $\bigOh(1)$  & Location and Spread, $\bigOh(1)$  & EM & Gaussian Approximation\tabularnewline
\hline 
TrueSkill2 & Continuous & Location and Spread, $\bigOh (1)$  & Location and Spread, $\bigOh (1)$  & EM & Gaussian
Approximation\tabularnewline
\hline
SMC & General & Full Distribution, $\bigOh (J)$  & Full Distribution, $\bigOh (J)$ \tablefootnote{The computational cost of SMC smoothing can be reduced from $\bigOh (J^2)$ to $\bigOh (J)$ using rejection sampling \citep{douc2011sequential} or MCMC \citep{dau2023backward} strategies.}  & EM & Monte Carlo Variance\tabularnewline
\hline 
Discrete & Discrete & Full Distribution, $\bigOh (S^{2})$  & Full Distribution, $\bigOh (S^{2})$ \tablefootnote{Full smoothing for a continuous-time fHMM has complexity $\bigOh (S^3)$, however for factorised random walk dynamics, marginal smoothing and gradient EM can be applied at cost $\bigOh (S^2)$, see A.6.3 in the supplementary material.}  & (Gradient) EM & N/A\tabularnewline
\hline 

\end{tabular}
% }
}
\end{table}

\section{Methods}\label{sec:Concrete-Procedures}

In this section, we turn to some concrete models for two-player competition as well as natural inference procedures. We here focus on the key components of the approaches, highlighting the probabilistic model used (for model-based methods) and the inference paradigm. For ease of presentation, our initial focus will be on pairwise comparison and sigmoidal observations with match outcomes $y \in \mathcal{Y} \subseteq \{\text{draw}, \text{home win}, \text{away win}\}$, but extensions to a multiplayer scenario and more general observation models are discussed in section \ref{subsec:extensions}. Detailed recursions for each approach can be found in Section A of the supplementary material. The presented methods alongside notable features are summarised in Table~\ref{tab:comparison}.

Potentially the simplest model for latent skills is a (static) Bradley-Terry model\footnote{Note that Bradley-Terry models are often (and indeed originally) described on an exponential scale i.e. in terms of $z = \exp(x) \in \real_+$.} \citep{Bradley1952}, wherein skills take values in  $\mathcal{X} = \real$ and (binary) match outcomes are modelled with the likelihood
\begin{align*}
    G^\text{BT} \left( y \mid x^{h}, x^{a}\right) = 
    \begin{cases}
        \sigmoid \left({x^{h} - x^{a}} \right) & \text{if } y = \mathrm{h},\\
        1 - \sigmoid \left({x^{h} - x^{a}} \right) & \text{if } y = \mathrm{a},
    \end{cases}
\end{align*}
where $\sigmoid: \real \to \left[ 0, 1 \right]$ is an increasing function which maps real values to normalised probabilities, such that $\sigmoid(x) + \sigmoid(-x) = 1$. The full Bradley-Terry model thus takes the form
\begin{align*}
    \mathbf{P} \left( x^{\left[ N \right]}, y_{1:K} \right)  = \prod_{i \in \left[ N \right]} m_0^{i} \left( x^{i} \right) \cdot \prod_{k=1}^K G^\text{BT} \left( y_k \mid x^{h(k)}, x^{a(k)}\right),
\end{align*}
with prior $m_0^{i}$. In practice, it is relatively common to neglect the prior and estimate the players' skills through pure maximum likelihood, see e.g. \cite{kiraly2017modelling}.

In contrast to the other models considered in this work, this Bradley-Terry model treats player skills as static in time. Therefore, as the `career' of each player progresses, our uncertainty over their skill level generally collapse to a point mass. We take the viewpoint that in many practical scenarios, this phenomenon is unrealistic, and so we advocate for models which explicitly model skills as varying dynamically in time. This leads naturally to the state-space model framework. 

\subsection{Elo}\label{subsec:Elo}
The Elo rating system \citep{Elo1978}, is a simple and transparent system for updating a database of player skill ratings as they partake in two player matches. Elo implicitly represents player skills as real-valued i.e. $\mathcal{X} = \real$, though typical presentations of Elo tend to eschew an explicit model. Skill estimates are updated incrementally in time according to the rule (for binary match outcomes)
\begin{align*}
    \begin{pmatrix}
        x_k^h \\ x_k^a
    \end{pmatrix}
    = 
    \begin{pmatrix}
        x_{k-1}^h + \gamma \cdot \left(\mathbb{I} \left[ y_k = \mathrm{h} \right] - \sigmoid \left( \frac{x_{k-1}^{h} - x_{k-1}^{a}}{s} \right) \right)
        \\
        x_{k-1}^a + \gamma \cdot \left(\mathbb{I} \left[ y_k = \mathrm{a} \right] - \sigmoid \left( \frac{x_{k-1}^{a} - x_{k-1}^{h}}{s} \right) \right)
    \end{pmatrix},
\end{align*}
where the sigmoid function is usually taken to be the logistic $\sigmoid_\mathrm{L} (x) = \left( 1+\exp(-x) \right)^{-1}$. Here $s$ is a scaling parameter and $\gamma$ is a learning rate parameter; these are typically set empirically for each competition, e.g. for Chess \citep{FIDE}, one takes $s = 400/\log(10)$ and $\gamma \in \{10, 20, 40\}$ depending on a player's level of experience. Note that rescaling $\left( \gamma, s \right)$ by a common factor leads to an essentially equivalent algorithm; it is thus mathematically convenient to work with $s = 1$ for purposes of identifiability. In practice, the ratio $\gamma / s$ is identifiable and carries an interpretation of the speed at which player skills vary on their intrinsic scale per unit of time. The Elo rating system can be generalised to give valid normalised prediction probabilities in the case of draws (i.e. ternary match outcomes) via the Elo-Davidson system \citep{Davidson1970, szczecinski2020understanding}, the recursions for which we outline in A.1 in the supplementary material.

% The Elo rating system has been used in a variety of settings, most famously in chess \citep{Elo1978} but also in a wide variety of other sports (see \cite{Stefani2011} for a review), for many of which it remains the official rating system used for e.g. seeding and matchmaking. 
The popularity of Elo arguably stems from its simplicity, as it can be well understood without a statistical background. Interestingly, it has also been shown to provide surprisingly hard-to-beat predictions in many cases \citep{Hvattum2010, Kovalchik2016}. However, we emphasise that Elo is not explicitly model-based, and this can make it difficult to extend to more complex scenarios, or to critique the assumptions by which it is underpinned.

\subsection{Glicko and Extended Kalman}\label{subsec:Glicko_Kalman}
It was noted in \cite{Glickman1999} that the Elo rating system is reminiscent of a Bradley-Terry model with a dynamic element. This observation led to the development of the Glicko rating system, which explicitly seeks to take into account time-varying uncertainty over each player's latent skill ratings, i.e. representing $\text{Filter}_k^i \approx \gauss \left( x_k^i \mid \mu_k^i, \sigma^{i \, 2}_k \right)$. This enriches the Elo approach by tracking both the location and spread of player skills at a given instant. The $\mathsf{Propagate}$ and $\mathsf{Assimilate}$ steps used in Glicko invite comparison to a (local, marginal) variant of the (Extended) Kalman filter (see e.g. Chapter 7 of \cite{sarkka2023bayesian}), a connection which was made formal in \cite{Ingram2021} and later \cite{Szczecinski2023}. Further details on Glicko can be found in \cite{Glickman1999}.

In Glicko, sports which permit draws are only heuristically permitted by treating them as `half-victories'; this implementation does not provide normalised prediction probabilities for sports with draws, which is undesirable. By adopting the Extended Kalman filter perspective, we can readily provide a principled approach to sports with draws by considering the following (non-linear) factorial state-space model (considered in \cite{Dangauthier2008} and \cite{Minka2018})
\begin{align}\label{specific_SSM}
    m_0 \left( x_0^i \right) = \gauss &\left(x_0^i \mid \mu_0, \sigma_0^2 \right), \qquad
    M_{t, t^\prime} \left(x^i_{t}, x^i_{t^\prime} \right) = \gauss \left( x^i_{t^\prime} \mid x^i_{t}, \tau^2 \cdot \left( t^\prime - t \right) \right), \\
    G_k \left( y_k \mid x^h, x^a \right)
    &=
    \begin{cases}
        \sigmoid \left(\frac{x^h - x^a + \epsilon}{s}\right) - \sigmoid\left(\frac{x^h - x^a - \epsilon}{s} \right) & \text{if } y_k = \text{draw}, \\
        \sigmoid \left(\frac{x^h - x^a - \epsilon}{s} \right) & \text{if } y_k = \mathrm{h},\\
        1-\sigmoid \left(\frac{x^h - x^a + \epsilon}{s} \right) & \text{if } y_k = \mathrm{a}.
    \end{cases}\nonumber
\end{align}

As with Elo, the logistic sigmoid function is typically used. Similarly, the parameters $\mu_0$, $\sigma_0$, $\tau$ and $s$ are not jointly identifiable, due to a translation and scaling equivariance. We break this symmetry by setting $\mu_0 = 0$, $s = 1$; note that in practice, implementations of Glicko will often use alternative numerical values in service of interpretability, comparability with Elo ratings, and so on.

The state-space model perspective elucidates the interpretation of the parameters $\sigma_0$, $\tau$ and $\epsilon$. The initial variance $\sigma_0^2$ controls the uncertainty over the skill rating of a new player entering the database, $\tau$ is a rate parameter that controls how quickly players' skill vary over time (note that $\tau=0$ recovers a static Bradley-Terry model) and $\epsilon$ is a draw parameter that dictates how common draws are for the given sport (for sports without draws, $\epsilon=0$).

The original presentation of Glicko in \cite{Glickman1999} observed that smoothing can be easily applied using the standard backward Kalman smoother \citep{sarkka2023bayesian}, which they used in service of `pure smoothing', rather than for parameter estimation. The parameters for Glicko ($\sigma_0$ and $\tau$) can be inferred by numerically minimising a cross-entropy-type loss function which contrasts predicted and realised match outcomes (see \cite{Glickman1999}, Section 4); this works reasonably in its own context, but does not necessarily scale well to more complex models with a larger parameter set. Adopting the framework described in Section \ref{subsec:ParamEst}, we determine maximum-likelihood estimators for the static parameters which scale gracefully to complex models and parameter sets. In particular, for \eqref{specific_SSM}, the convenient Gaussian form of the smoothing approximation means the maximisation step of EM can be carried out analytically for $\sigma_0$ and $\tau$, and efficiently numerically for $\epsilon$ \citep{ghq}.

\subsection{TrueSkill and Expectation Propagation/Moment-Matching}\label{subsec:TrueSkill}

It was noted in \cite{Herbrich2006} that if we instead choose the sigmoid function for a static Bradley-Terry model to be the inverse probit function $\sigmoid_{\mathrm{IP}} \left( x \right) = \Phi \left( x \right)$ (where $\Phi$ is the CDF of a standard Gaussian), then certain integrals of interest become analytically tractable. In particular, we can use the identity $\int \gauss(z \mid \mu, \sigma^2) \cdot \Phi(z) \, \mathrm{d}z = \Phi ({\mu} \slash {\sqrt{1 + \sigma^2}})$ to analytically calculate the marginal filtering means and variances of the non-Gaussian joint filtering distribution $\text{Filter}^{h,a}_k$. This naturally motivates the moment-matching approach of \cite{Herbrich2006}, wherein an approximate factorial filtering distribution $\text{Filter}^{h,a}_k \approx \text{Filter}^{h}_k \cdot \text{Filter}^{a}_k$ can be defined by simply extracting the marginal means and variances. This moment-matching strategy can be seen as a specific instance of \textit{assumed density filtering} (see e.g. Chapter 1 of \cite{minka2001family} and references therein) or its more general cousin, Expectation Propagation \citep{minka2013expectation}. We provide some details on this connection in Section C in the supplementary material. One can also reasonably consider applying other approximate filtering strategies based on Gaussian principles (e.g. Unscented Kalman Filter \citep{julier2004unscented}, Ensemble Kalman Filter \citep{evensen2009data}, etc.) to the same model class; we do not explore this further here.

In the original TrueSkill \citep{Herbrich2006}, this procedure was applied as an approximate inference procedure in the static Bradley-Terry model. In the follow-up works TrueSkillThroughTime \citep{Dangauthier2008} and TrueSkill2 \citep{Minka2018}, the model was extended to allow the latent skills to vary over time. The resulting procedure is a treatment of the state-space model in \eqref{specific_SSM}, where the filtering distributions are formed by i) applying the decoupling approximation, and ii) assimilating observations with predictions through the aforementioned moment-matching procedure. Smoothing is handled analogously to the Glicko and Extended Kalman settings, i.e. by running the Kalman smoother backwards from the terminal time. TrueSkill2 \citep{Minka2018} applies a gradient-based version of the parameter estimation techniques presented in Section \ref{subsec:ParamEst}, although there it is not presented explicitly in the state-space model context.

We note that the above description is TrueSkill \citep{Herbrich2006, Minka2018} in its most basic form. The TrueSkill approach has been successfully applied in more complex settings, notably for online multiplayer games \citep{Minka2018}.

\subsection{Sequential Monte Carlo}\label{subsec:SMC}

The preceding strategies all hinge on the availability of a suitable parametric family for approximating the relevant probability distributions, enabling explicit computations and general ease of construction. This is counter-balanced by the limited flexibility of parametric approximations, where even in the presence of an increased computational budget, it is not always clear how to obtain improved estimation performance. This can sometimes be ameliorated by the use of nonparametric approximations, in the form of particle methods like sequential Monte Carlo (SMC). SMC can be applied to any state-space model for which we can i) simulate from both the initial distribution $m_0$ and the Markovian dynamics $M_{t, t^\prime}$ and ii) evaluate the likelihood $G_k$. Naturally, in this work, it is of particular interest to consider the application of SMC to the (factorial) state-space model in \eqref{specific_SSM}. It is out of the scope of this paper to review available variants of SMC. We instead prioritize conciseness by concentrating on the most widely-used instance of SMC, the bootstrap particle filter. 

%This can sometimes be ameliorated by the use of nonparametric approximations, in the form of particle methods. In particular, we will consider a sequential Monte Carlo (SMC) strategy based on importance sampling. 

%SMC encompasses a diverse range of algorithms which exhibit variations through their choice of proposal distribution and resampling scheme. It is out of the scope of this paper to review available variants of SMC. We instead prioritize conciseness by concentrating on the most widely-used instance of SMC, the bootstrap particle filter. 

SMC filtering maintains a (potentially weighted) particle approximation to the filtering distributions which in our context has an additional factorial approximation, reminiscent of a `local' or `blocked' SMC approach \citep{rebeschini2015can},
$\mathrm{Filter}_k^i(x_k^i) \approx \sum_{j \in \left[ J \right]} w_{k}^{i \,j} \cdot \delta (x_{k}^{i} \mid x_{k}^{i\,j})$,
where $\delta(x \mid y)$ is a Dirac measure in $x$ at point $y$ and $j \in \left[ J \right]$ indexes particles.

% %
% \begin{align*}
%     \mathrm{Filter}_k^i(x_k^i) \approx \sum_{j \in \left[ J \right]} w_{k}^{i \,j} \cdot \delta \left(x_{k}^{i} \mid x_{k}^{i\,j}\right),
% \end{align*}
% %
% where $\delta(x \mid y)$ is a Dirac measure in $x$ at point $y$ and $j \in \left[ J \right]$ indexes particles. 

The bootstrap particle filter then executes $\mathsf{Propagate}$ by simply simulating from the dynamics $M_{k, k+1}$ to provide a new particle approximation to $\text{Predict}_{k+1|k}^i$. Before applying the $\mathsf{Assimilate}$ step, the distributions $\text{Predict}_{k+1|k}^h$ and $\text{Predict}_{k+1|k}^a$ are paired together to form a joint distribution $\text{Predict}_{k+1|k}^{h, a}$. The $\mathsf{Assimilate}$ step then consists of a reweighting step and a resampling step (to carry forward only high-probability particles) resulting in a joint weighted particle approximation to $\text{Filter}_k^{h, a}$. The factorial approximation can then be regained by a simple $\mathsf{Marginalise}$ operation which unpairs the joint particles. Note that the factorial approximation is nonstandard in an SMC context, and is adopted here as a natural means of avoiding the curse of dimensionality that affects SMC \citep{rebeschini2015can}.

Smoothing can be applied using a similar iterative importance sampling approach \citep{godsill2004monte} that sweeps backwards for $k= K {-} 1, \dots, 0$ recycling the filtering approximations $\text{Filter}_{k}^{i}$ into joint smoothing approximations $\text{Smooth}_{0 : K | K}^{i}$. This procedure is (embarrassingly) parallel in both the number of players $N$ and the number of particles $J$, given the filtering approximations; see \cite{finke2017approximate} for a similar scenario. Parameter estimation can also be achieved by applying general-purpose expectation-maximisation or gradient ascent techniques from Section~\ref{subsec:ParamEst}, where the integrals (\ref{eq:em_q}-\ref{eq:em_grad}) are approximated using particle approximations to the smoothing law. Full details can be found in Section A.5 of the supplementary material and \cite{Chopin2020} provides a thorough review of the field.

\subsection{Finite State-Space}\label{subsec:fHMM}

From a modelling perspective, it is conceptually simple to consider skills which take values in a finite state-space. Recalling the general model formulated in \eqref{eq:joint-lik}, by choosing $\mathcal{X} = \left[ S \right]$ for some $S \in \nat$, one obtains a factorial hidden Markov model.
%By varying $S$, one has the freedom to adapt the flexibility of the model to the richness of the data. 

In this finite state-space, it is natural to model the skills of player $i$ as evolving according to a continuous time Markov jump process. In the time-homogeneous case, such processes can be specified in terms of their so-called “generator matrix” $Q_S$, an $S \times S$ matrix which encodes the rates at which the player's skill level moves up and down, and is typically sparse. Given such a matrix, it is typically straightforward to construct the corresponding transition kernels $M_{t, t^\prime}$ at a cost of $\bigOh (S^3)$ by diagonalisation and matrix exponentiation. In many settings, such as filtering, one is not interested in the transition matrix $M_{t, t^\prime}$ itself but rather its action on probability vectors. In this case, given appropriate pre-computations, the cost can often be controlled at the much lower $\bigOh (S^2)$. We provide further details in Section A.6 of the supplementary material.

% In these models, the likelihood $G_k$ then takes the form of an $S \times S \times \mathbf{card}(\mathcal{Y})$ array, representing the probabilities of observing a certain outcome given a certain pair of player skills. For modelling coherence, it is natural that this array satisfies certain monotonicity constraints, so that i.e. if a player's skill level increases, then they should become more likely to win matches.\sp{Y not nec finite}

In the context of skill ratings with pairwise comparisons, we can then consider the following fHMM inspired by \eqref{specific_SSM}. Writing $Q_S$ for the generator matrix of the continuous-time random walk with reflection on $\left[ S \right]$ (see Section B of the supplementary material) and using $\exp$ for the matrix exponential, we can define
\begin{align}\label{specific_HMM}
    m_0 &= \nu \cdot M_{0,\sigma_d}, \qquad M_{t, t^\prime} = \exp \left( \tau_{d} \cdot \left(t^{'}-t\right) \cdot Q_S \right), \\
    G_k \left( y_k \mid x^h, x^a \right)
    &=
    \begin{cases}
        \sigmoid \left(\frac{x^h - x^a + \epsilon_d}{s_d}\right) - \sigmoid\left(\frac{x^h - x^a - \epsilon_d}{s_d} \right) & \text{if } y_k = \text{draw}, \\
        \sigmoid \left(\frac{x^h - x^a - \epsilon_d}{s_d} \right) & \text{if } y_k = \mathrm{h},\\
        1-\sigmoid \left(\frac{x^h - x^a + \epsilon_d}{s_d} \right) & \text{if } y_k = \mathrm{a},
    \end{cases} \nonumber
\end{align}
with subscript $d$ used to emphasise the connection to the discrete model. 

Here $\nu$ is a probability vector whose mass is concentrated on the median state(s) $\left\{ \left\lfloor \frac{S}{2} \right\rfloor ,\left\lceil \frac{S}{2} \right\rceil \right\}$, so that $m_0$ resembles a (discrete) centred Gaussian law with standard deviation of order $\sigma_d$ (for $\sigma_d \ll \sqrt{S}$). As one takes $\sigma_d \to \infty$, this converges towards the uniform distribution on $\left[ S \right]$. Similarly to \eqref{specific_SSM}, for the dynamical model, we have a rate parameter $\tau_d \in \real^+$ which controls how quickly skill ratings vary over time, and for the observation model, we have a scaling parameter $s_d  \in \real^+$ and a draw propensity parameter $\epsilon_d  \in \real^+$; this can be set as $\epsilon_d=0$ for sports without draws.

% For inference, we can follow the procedure described in \cite{rimella2022exploiting}, which is well-suited to our highly-localised observation model. For filtering, at time $k$ we can perform the steps described in Section \ref{sec:pairwise_updates}, which in the fHMM scenario can be implemented in closed form via simple linear algebra operations, representing the filtering laws as normalised probability (row) vectors of length $S$. 

% Computationally-efficient smoothing follows from observing that both the filtering distributions and the transition kernel factorise across players. That is, the equations in Section \ref{sec:smoothing} can be applied exactly through matrix multiplications and independently on each player.

Computationally efficient filtering and smoothing follow from observing that both the filtering distributions and the transition kernel factorise across players, as in \cite{rimella2022exploiting}. That is, the equations in Section \ref{sec:smoothing} can be implemented in closed form via simple linear algebra operations.

Parameter estimation can be performed through the EM algorithm. For parameters not associated with the dynamical model, the $\mathbf{Q}$ function can be formed using only the marginal smoothing laws $\mathrm{Smooth}_{k \mid K}^i$, which are available at a cost of $\bigOh(S^2)$. EM updates for the dynamical parameter $\tau_d$ instead require joint laws of the form $\mathrm{Smooth}_{k, k + 1 \mid K}^i$, which are more costly to assemble; we thus opt to instead update $\tau_d$ by a cheaper gradient ascent step.

\subsection{Model Extensions} \label{subsec:extensions}

The generality and flexibility of the state-space approach permits the modelling of more sophisticated data-generating processes with minimal or no alteration to the inference procedures. In simple cases, this might involve simply refining the static parameters. For instance, we could learn distinct $s_{\text{3-set}} > s_{\text{5-set}}$ to model the additional randomness associated with the shorter tennis format. Similarly, we could separate $\tau_\text{off-season}$ and $\tau_\text{in-season}$ to consider different levels of skills fluctuation during in- and off-season periods. We could even consider players' specific parameters, for example, learning player-specific initial distribution parameters or $\tau^i$ to allow player skills to evolve differently. %This could be achieved with minimal modification to the EM algorithm.

We now describe some more involved model modifications including how popular existing models can be re-framed within the state-space model framework.

%The generality of the factorial state-space approach permits great flexibility to model more sophisticated data generating processes with minimal or no modification to the inference procedures. In simple cases, this could be enhancing the static parameters. For example, we could use distinct $s_{\text{3-set}} > s_{\text{5-set}}$ to model the additional randomness associated with the shorter tennis format or perhaps separate $\tau_\text{off-season}$ and $\tau_\text{in-season}$. We now describe some more involved model modifications including how popular existing models can be re-framed within the factorial state-space model framework.
%\sd{Add comment on non-identical dynamnics and dynamics parameters}

\subsubsection*{Stationary Dynamics: Ornstein-Uhlenbeck}

A natural modification of the dynamics in \eqref{specific_SSM} (which we use following the works of Glicko \citep{Glickman1999} and TrueSkill \citep{Dangauthier2008}) would be to replace the Brownian motion skill evolution with that of an Ornstein-Uhlenbeck process $$M_{t, t^\prime} \left(x^i_{t}, x^i_{t^\prime} \right) = \gauss \left( x^i_{t^\prime} \mid x^i_{t} \cdot e^{-\tau (t^\prime - t)} + \mu_0 \cdot (1 - e^{-\tau (t^\prime - t)}),  \sigma_0^2 \cdot (1 - e^{- 2\tau (t^\prime - t)}) \right).$$ 
Now each player's skill is mean reverting towards $\mu_0$; this behaviour may be more desirable depending on the sport at hand. Contrarily, we may believe that the average ability of players in a sport has some underlying trend over time which may be better captured by the Brownian dynamics. Note that stationary dynamics are already used in \eqref{specific_HMM} as the finite-space dynamical process reverts to the uniform distribution. 
% \sd{Mention discussion of needing a $\mu_0$ for each player and stable long-time variance. Perhaps defer in-depth discussion to appendix.}

\subsubsection*{Points-based Match Outcomes: Bivariate Poisson}

% The popular \cite{dixon1997modelling} model uses an inflated Poisson likelihood for the number of home and away goals in a football (or similar points scoring) match, meaning $y_k = (y_k^h, y_k^a) \in \nat^2$ . Furthermore the players' skill is considered as bivariate with $x^i = (x^{\text{att}, i}, x^{\text{def}, i})$ and where $x^{\text{att}, i}, x^{\text{def}, i}$ model the attack and defence strength of player $i$ respectively. The observation model is given by
% $$
% G_k (y_k \mid x^h, x^a) 
%     = e^{-(\lambda_1 + \lambda_2)} \frac{\lambda_1^{y_k^h}}{y_k^h!} \frac{\lambda_2^{y_k^a}}{y_k^a!} \rho \left(y_k, \lambda_1, \lambda_2, \iota\right),
% $$
% where $\lambda_1 = \alpha{x^{\text{att}, h}x^{\text{def}, a}}$, $\lambda_2 ={x^{\text{att}, a}x^{\text{def}, h}}$ and the function $\rho$ control the level of inflation of certain outcomes whilst $\alpha$ and $\iota$ are static parameters to be learnt on a constrained space (see \cite{dixon1997modelling} and supplementary material for details), alongside $\sigma_0^\text{att}$, $\sigma_0^\text{def}$ and $\tau$ (with fixed $\mu_0^\text{att} = \mu_0^\text{def} = 0$ for identifiability).

The popular \cite{dixon1997modelling} model uses an inflated Poisson likelihood to model the home and away goals scored in a football (or similar points scoring) match, meaning $y_k = (y_k^h, y_k^a) \in \nat^2$. Each player's skill is considered as bivariate with $x^i = (x^{\text{att}, i}, x^{\text{def}, i}) \in \real^2$ and where $x^{\text{att}, i}, x^{\text{def}, i}$ model the attack and defence strength of player $i$ respectively.
More generally, \cite{karlis2003analysis} use a bivariate Poisson likelihood
\begin{equation}\label{bivariate_poisson}
G_k (y_k \mid x^h, x^a) 
    = e^{-(\lambda_1 + \lambda_2+\lambda_3)} \frac{\lambda_1^{y_k^h}}{y_k^h!} \frac{\lambda_2^{y_k^a}}{y_k^a!} \sum_{k=0}^{\min\{y_k^h, y_k^a\}} \binom{y_k^h}{k} \binom{y_k^a}{k} k! \left ( \frac{\lambda_3}{\lambda_1 \lambda_2}\right )^k,
\end{equation}
where $\lambda_1 = \exp(\upalpha^h + {x^{\text{att}, h} - x^{\text{def}, a}})$, $\lambda_2 = \exp(\upalpha^a + {x^{\text{att}, a} - x^{\text{def}, h}})$, $\lambda_3 =\exp(\beta)$ and $\upalpha^h, \upalpha^a, \beta \in \real$ are all static parameters (to accompany the static parameters for the initialisation and dynamics). 

These likelihoods can be easily integrated within a state-space model to infer how different teams' attacking and defensive strengths vary over time. The factorial assumption remains and the players' multivariate skills evolve independently preserving the form of $\mathsf{Propagate}$. On the other hand, $\mathsf{Assimilate}$ should be modified to update jointly $x^{\text{att}, h}, x^{\text{def}, h}, x^{\text{att}, a}, x^{\text{def}, a}$, which is then followed by $\mathsf{Marginalise}$ to regain the factorial approximation.

Such model extensions can lead to new inferences of interest where one might want to query the existence of covariates representing external factors \citep{karlis2003analysis}. As a concrete example, one could examine the presence of a `home advantage' factor in football by testing the hypothesis $\upalpha^h > \upalpha^a$ against $\upalpha^h = \upalpha^a$. 

\subsubsection*{Multiplayer Extension: Plackett-Luce}
% So far we have primarily focused on pairwise comparisons, where matches are assumed to be between two players. We can extend this formulation to a set of players $C(k)$ resulting in $G_{k} \left( y_{k} \mid x_{{k}}^{C(k)}\right)$, which can be of broad use e.g. in multiplayer games \cite{Joshy2024}, movie rankings \cite{guiver2009bayesian} and consumer behaviour \cite{Luce1959individual}.  Steps 1 to 5 from Section \ref{sec:pairwise_updates} follow in the same way and precisely require to: retrieve the latest match times of all the players involved in the match and the corresponding filtering distributions, compute predictive distributions for them, assimilate the new result and then marginalise back to a factorial approximation. In particular, the $\mathsf{Assimilate}$ procedure in step 4 requires mapping to a joint distribution on all players in $C(k)$ and obviously the $\mathsf{Marginalise}$ operation in step 5 is applied on all players as well.  

We have thus far focused on pairwise comparisons, where matches are assumed to be between two players. We can extend our formulation to a collection of players which can be of broad use e.g. in multiplayer games \cite{Joshy2024}, movie rankings \cite{guiver2009bayesian} and consumer behaviour \cite{Luce1959individual}. All steps in Section \ref{sec:pairwise_updates} remain conceptually the same, with the $\mathsf{Assimilate}$ and $\mathsf{Marginalise}$ steps now acting on an enlarged intermediate joint distribution.

Specifically, we can consider the model of \cite{plackett1975analysis, Luce1959individual}, where a \textit{judge} ranks $\textsf{M}$ players competing in match $k$. The observed data is $y_k=(y^1_k, \dots , y_k^\textsf{M})$ where each $y_k^\textsf{m}$ uniquely indexes one of the $\textsf{M}$ player indices, e.g. $y_k^\textsf{3} = 26$ indicates the judge ranked player 26 in 3rd place. The likelihood \citep{guiver2009bayesian} is
\begin{align*}
    G_k\left(y_k^{1:\mathsf{M}} \mid x^{[\mathsf{M}]}\right)
    = \prod_{\mathsf{m} = 1}^\mathsf{M} \frac{
    \exp({x^{y_k^{\mathsf{m}}}})
    }{\exp({x^{y_k^{\mathsf{m}}}}) + \dots + \exp({x^{y_k^{\mathsf{M}}}})
    }.
\end{align*}
As this likelihood applies to $\textsf{M}$ players simultaneously (as opposed to just two previously) care must be taken with the dimensionality of the assimilation for SMC and discrete inference. In principle, the extended Kalman approach does not suffer in the same way, although the accuracy of the Taylor approximation may need to be validated.

\section{Experiments\label{sec:Experiments}}

We now turn to the task of accurately quantifying sporting skills with some real-life data sets. We consider three sports; Women's Tennis Association (WTA) results (noting that all matches have the same 3-set format), football data for the English Premier League (EPL) (plus international data for Fig.~\ref{fig:argentina}) and professional (classical format) chess matches. In all cases, the research question is to use the results of the matches (and potentially additional data, e.g. the margin of victory) to infer time-varying skill levels of all competitors alongside uncertainties in these skill estimates. We then use these skill levels to provide match predictions, whose accuracy we assess.

This section is structured to replicate a realistic workflow. We start with an exploratory analysis with some trial static parameters, testing against basic coherence checks on how we expect the latent skills to behave. We then turn to parameter estimation and learn the static parameters from historical data. Finally, we describe and analyse how filtering and smoothing can be utilised for online decision-making and historical evaluation respectively. At each stage of the workflow, we compare and highlight similarities or differences across sports and modelling or inference approaches as appropriate (but not exhaustively). %We consider all (dynamic) methods discussed in Section~\ref{sec:Concrete-Procedures}.

Specific details and parameter specifications for the experiments can be found in Section D in the supplementary material. A \texttt{python} package permitting easy application of the discussed techniques, and code to replicate all of the following simulations, can be found at \href{https://github.com/SamDuffield/abile}{github.com/SamDuffield/abile}.

\subsection{Exploratory Analysis}

The first step in any statistical procedure is to explore the model with some preliminary (perhaps arbitrarily chosen) static parameters. The goal here is not a thorough evaluation of skill ratings, but rather to assess our prior intuitions.

\begin{figure}[h!]
    \centering
    \begin{minipage}{0.47\linewidth}
        \centering
        \includegraphics[width=\textwidth]{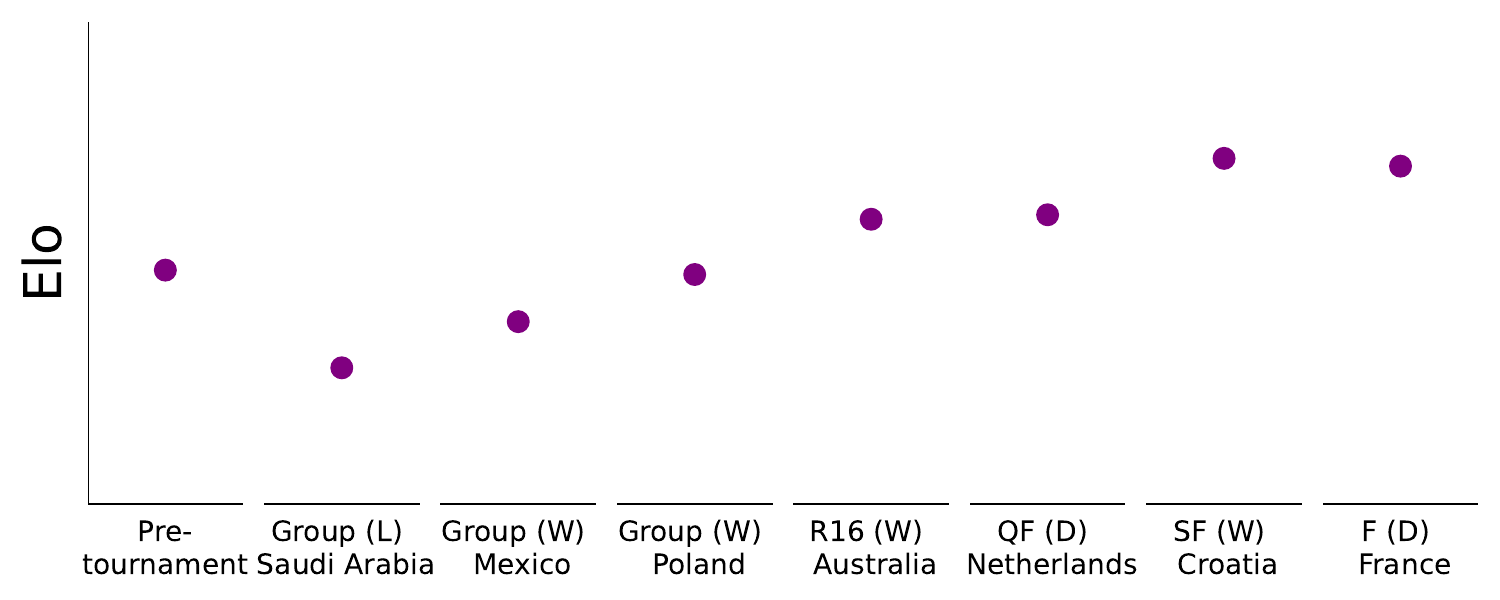}
    \end{minipage}
    \begin{minipage}{0.47\linewidth}
        \centering
        \includegraphics[width=\textwidth]{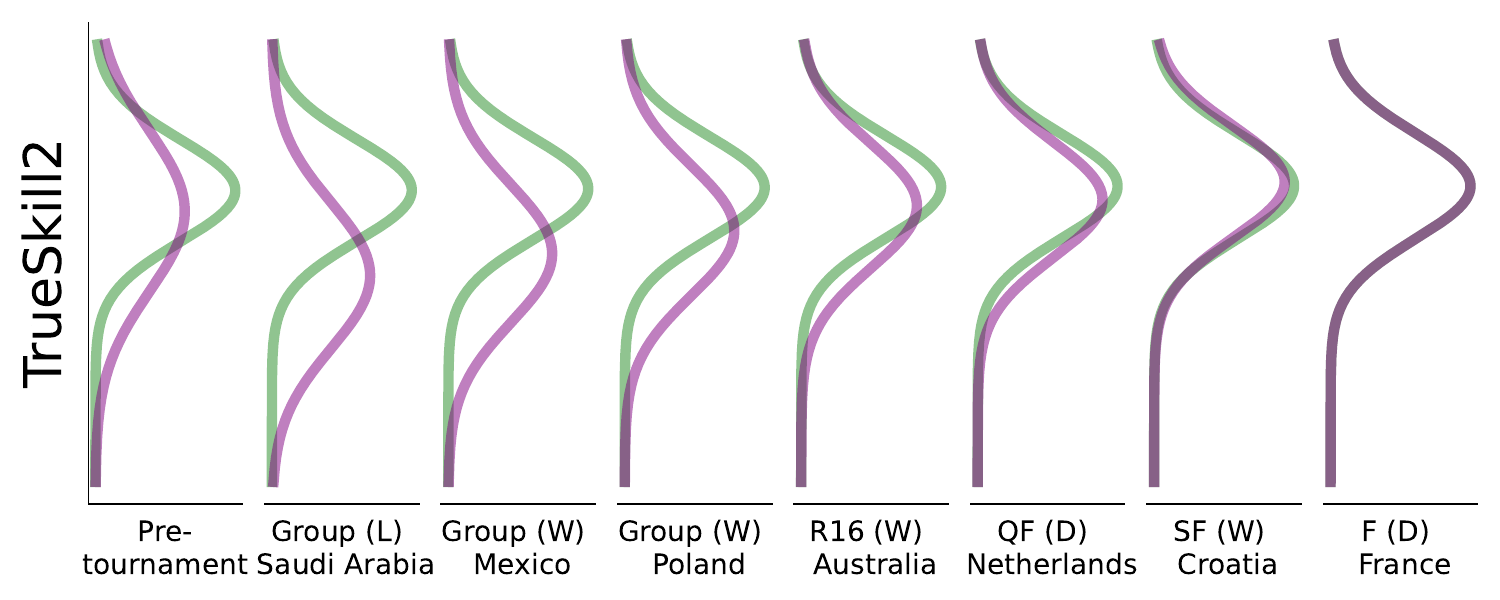}
    \end{minipage}
    \begin{minipage}{0.47\linewidth}
        \centering
        \includegraphics[width=\textwidth]{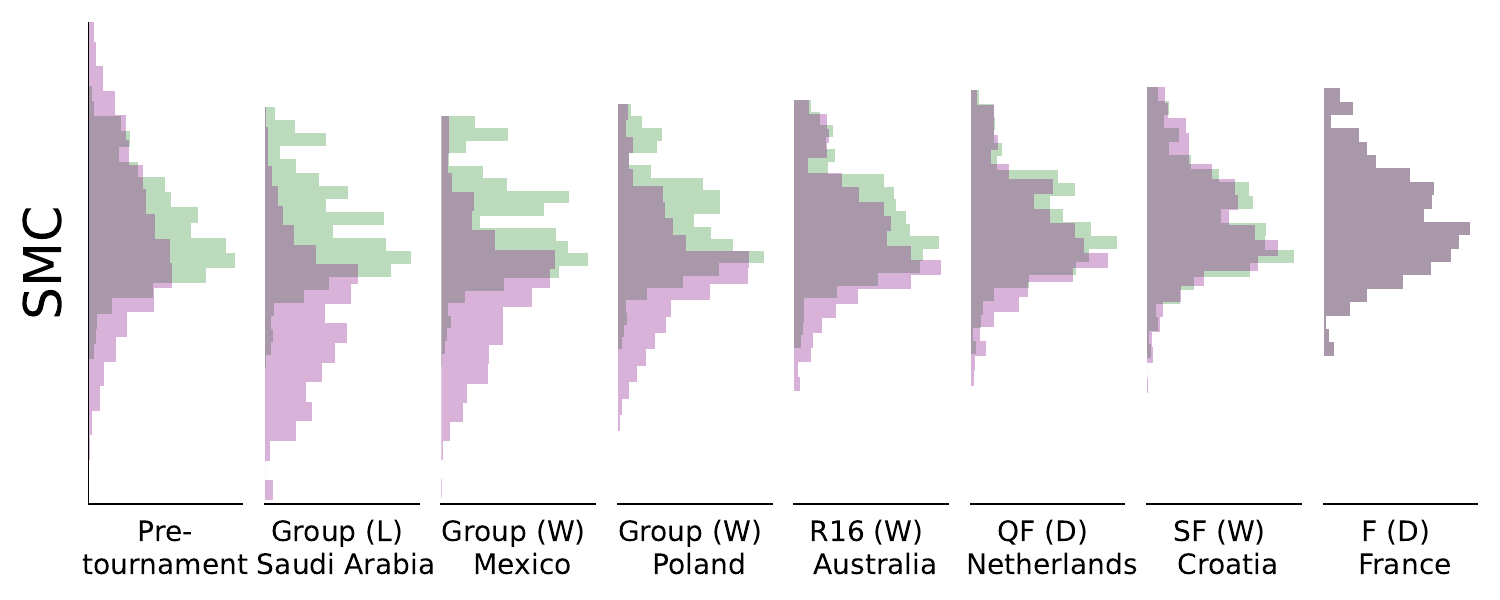}
    \end{minipage}
    \begin{minipage}{0.47\linewidth}
        \centering
        \includegraphics[width=\textwidth]{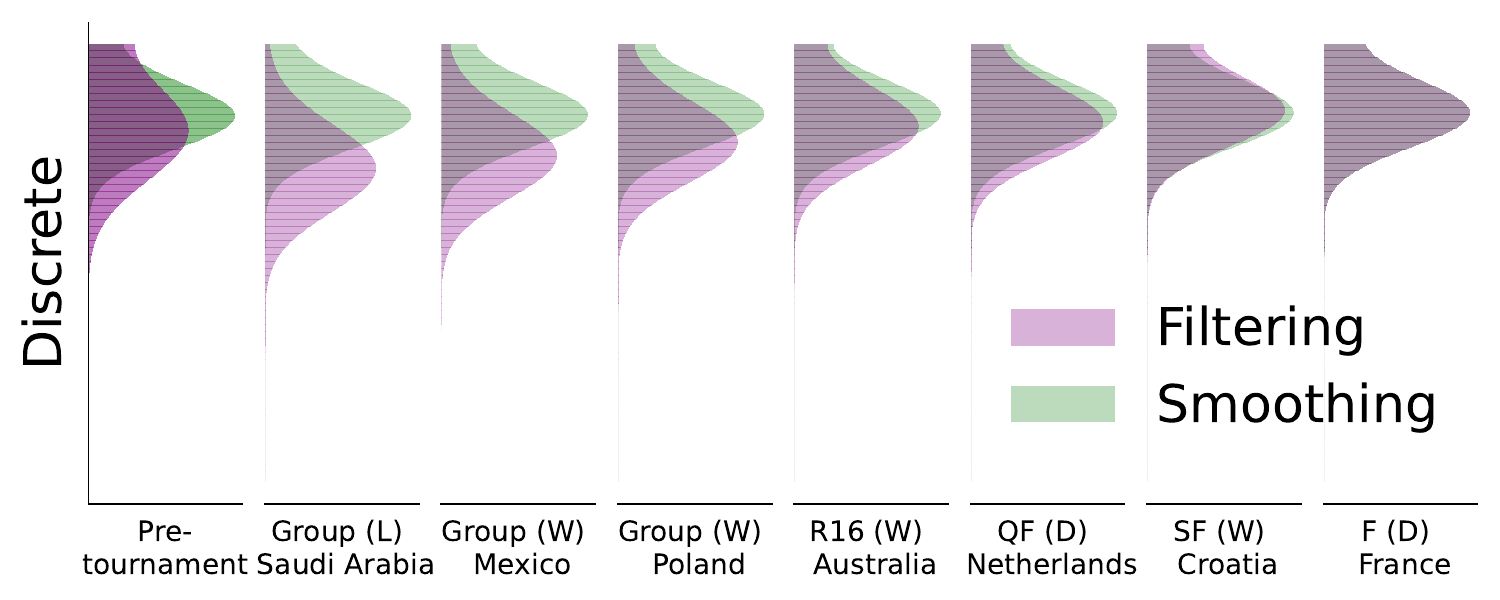}
    \end{minipage}
    \caption{Visualisation of the different skill representations for Argentina's 2023 FIFA World Cup triumph. Each y-axis represents the skill-rating scale for the different approaches.% (only SMC and TrueSkill2 share the same model and therefore also y-axis scaling).
    }
    \label{fig:argentina}
\end{figure}

In Fig.~\ref{fig:argentina}, we depict Argentina's 2023 football World Cup in terms of the evolution of their skill rating distribution with arbitrarily chosen static parameters (we run on international matches in 2020-2022, with only Argentina's post-match 2022 World Cup ratings displayed). We also take this opportunity to highlight the different approaches to encoding a probability distribution over skill ratings. We first see that Elo stores skill ratings as a point estimate without any uncertainty quantification and that it is a purely forward-based approach without any ability to update skill ratings based on future match results (i.e. smoothing). In contrast, the three model-based SSM approaches encode a (more informative) distribution over skill ratings. In the case of TrueSkill2 (Glicko and Extended Kalman would appear similar and are therefore omitted) a simple location and spread (Gaussian) distribution is used, whereas SMC and the discrete model can encode more complex distributions.

In terms of verifying our intuitions, we can see that Argentine victories increase their skill ratings, draws have little influence and defeats decrease the ratings. Particularly poignant is Argentina's defeat to Saudi Arabia (a low-ranked side according to both FIFA's official rankings and our computed filters) which in all approaches resulted in a sharp decrease in Argentina's estimated skill rating.

We can also draw insights from the smoothing distributions. We first note that at the final match in the dataset, the filtering and smoothing distributions match exactly, by definition. We also see that the smoothing distributions show less uncertainty than the filtering distributions; this matches our intuition since smoothing $\mathbf{P}(x_k \mid y_{1:K})$ has access to more data than filtering $\mathbf{P}(x_k \mid y_{1:k})$. We finally observe that the smoothing distributions are much less reactive to individual results and instead track a \textit{smooth} trajectory of the team's skill rating over time.

\subsection{Parameter Estimation}

Having verified informally that the algorithms are behaving sensibly, we turn to apply the techniques discussed in Section \ref{subsec:ParamEst} to learn the static parameters from historical data in an offline setting. Our goal is to maximise the log-likelihood $\log \mathbf{P}(y_{1:K} \mid \theta)$.

\begin{figure}[h!]
    \centering
    \includegraphics[width=0.8\linewidth]{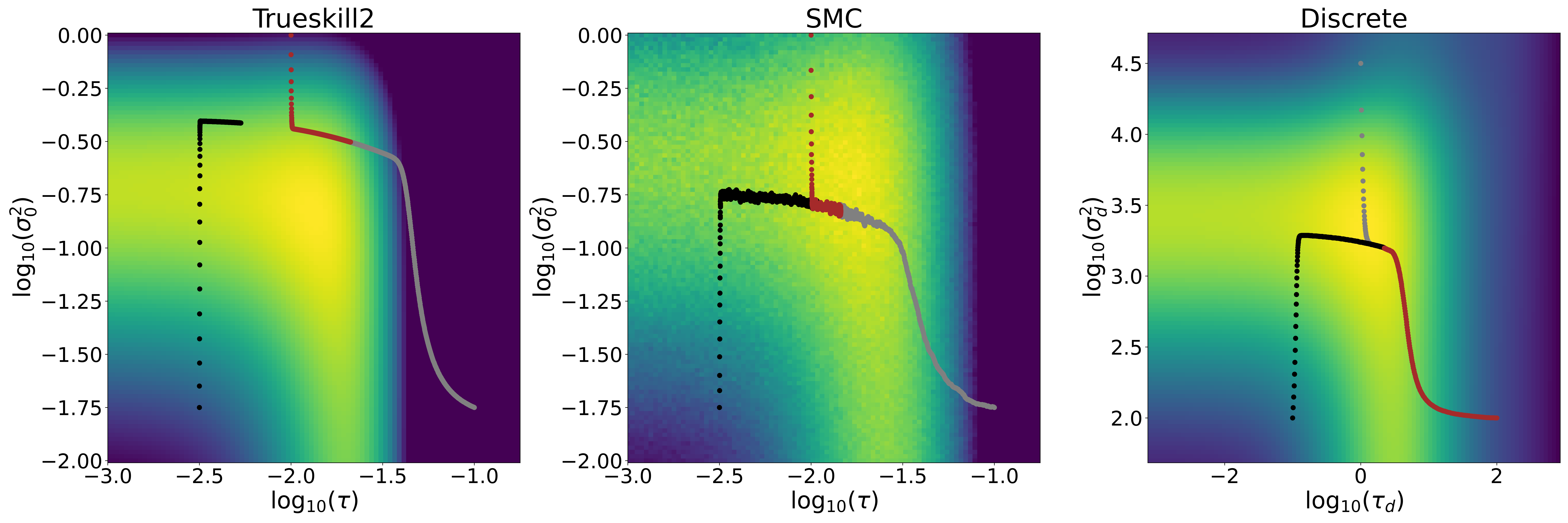}
    \caption{Log-likelihood grid and parameter estimation for WTA tennis data.
    % Note that TrueSkill2 and SMC share the same model.
    }
    \label{fig:train_tennis}
\end{figure}

We initially consider the WTA tennis dataset, which we train on the years 2019-2021 and leave 2022 as a test set for later. Draws do not occur in tennis, and therefore we can set $\epsilon = \epsilon_d = 0$, leaving only two parameters to tune for each approach. This two-dimensional optimisation landscape of the log-likelihood can readily be visualised, as in Fig.~\ref{fig:train_tennis}. Indeed, as the static parameter is only two-dimensional the optimisation could be applied using a grid search (as is indeed the most natural option for Elo and Glicko). Furthermore, a filtering sweep provides an estimate of the optimisation objective $\log \mathbf{P}(y_{1:K} \mid \theta)$, and therefore the grid search can be applied directly without running a smoothing routine. For more complex models and datasets, higher dimensional static parameters are inevitable, and a grid search will quickly become prohibitive. We therefore apply iterative expectation-maximisation to the tennis data in Fig.~\ref{fig:train_tennis} to investigate some of the properties and differences between approaches, indicative of parameter estimation in more complex situations.

For the three approaches considered (we again omit the Extended Kalman approach due to its similarity with TrueSkill2 in all steps beyond filtering), we display 1000 expectation-maximisation iterations starting from three different initialisations.

The TrueSkill2 and SMC approaches share the same model and differ only through their respective Gaussian and particle-based approximations of the skill distributions. The Gaussian approximation induces a significant bias, whereas the particle approximation is asymptotically unbiased, but induces Monte Carlo variance. We can see that the bias from the Gaussian approximation contorts the optimisation landscape for TrueSkill2 relative to SMC, whose fuzzy landscape is due to the stochastic nature of the algorithm. We see that the additional bias from the TrueSkill2 approach results in the EM trajectory evading the global optimum; by contrast, this optimum is successfully identified by both the SMC and discrete approaches, which do not exhibit any systematic bias beyond the factorial approximation.

\subsection{Filtering (for online decision-making)}

Now that we have a principled choice for our static parameters, we can apply the methods to online data. In Table~\ref{tab:num_comparison}, we use static parameters (trained using grid search for Elo and Glicko, and EM for the remaining, model-based approaches). Recall that in the case of sports with draws, Glicko does not provide the normalised outcome predictions required for log-likelihood-based assessment.

\begin{table}[h!]
\caption{\label{tab:num_comparison}Average negative log-likelihood (low is good) for match outcomes across a variety of sports. In each case, the training period was 3 years and the test period was the subsequent year. Note the draw percentages were 0\% for tennis, 22\% for football and 65\% for chess. Note that although the bivariate Poisson likelihood models home and away goals, we here report the statistics for predicting draw, home win or away win, which are accessible as a byproduct of the bivariate likelihood.}
\centering
% \fbox{
\small
\begin{tabular}{|l|ll|ll|ll}
\hline
\multirow{2}{*}{Method} & \multicolumn{2}{c|}{Tennis (WTA)} & \multicolumn{2}{c|}{Football (EPL)} & \multicolumn{2}{c|}{Chess}   \\
 % $ -\frac1K \log \mathbf{P}(y_{1:K} \mid \theta)$ 
 & Train           & Test            & Train            & Test             & Train         & \multicolumn{1}{c|}{Test}           \\
 \hline 
 \hline
Elo-Davidson    & 0.640           & 0.636           & 1.000            & 0.973           & 0.802          & \multicolumn{1}{l|}{1.001}          \\ \hline
Glicko           & 0.640           & 0.636           & \multicolumn{1}{c}{-}                & \multicolumn{1}{c|}{-}               & \multicolumn{1}{c}{-}              & \multicolumn{1}{c|}{-}            \\ \hline
Extended Kalman & 0.640           & \textbf{0.635}  & 0.988            & 0.965           & \textbf{0.801} & \multicolumn{1}{l|}{\textbf{0.972}  }          \\ \hline
TrueSkill2       & 0.650           & 0.668           & 1.006            & \textbf{0.961}  & 0.802          & \multicolumn{1}{l|}{0.978 }         \\ \hline
SMC             & 0.640           & 0.639           & 0.988            & 0.962           & \textbf{0.801} & \multicolumn{1}{l|}{0.974 }        \\  \hline
Discrete        & \textbf{0.639}  & 0.636           & \textbf{0.987}   & \textbf{0.961}  & \textbf{0.801} & \multicolumn{1}{l|}{0.976} \\ 
\hline
% \hline
\noalign{\vskip 1mm}
\cline{1-5}
\multicolumn{3}{|l|}{Bivariate Poisson - Extended Kalman} & \textbf{0.975} & 0.954  &   &   \\
\cline{1-5}
\multicolumn{3}{|l|}{Bivariate Poisson - SMC} & 0.978 & \textbf{0.950}  &   &   \\
\cline{1-5}
% \multicolumn{3}{|l|}{Bivariate Poisson - Discrete (M=10)} & 0.9844 & 0.9539  &   &   \\
% \cline{1-5}
% \multicolumn{3}{|l|}{Bivariate Poisson - Discrete (M=20)} & 0.9844 & 0.9536  &   &   \\
% \cline{1-5}
\multicolumn{3}{|l|}{Bivariate Poisson - Discrete} & 0.984 & 0.954  &   &   \\
\cline{1-5}
\end{tabular}
% }
\end{table}

For the tennis data, we notice that all models broadly perform quite similarly, with the exception of TrueSkill2; we suspect that this stems from the parameter estimation optimisation issues discussed above. The tennis task without draws represents a simpler binary prediction problem, and it is therefore not surprising that (with principled parameter estimation) predictive performance saturates. For the more difficult tasks of Football and Chess (where draws do occur), we see that the model-based approaches equipped with uncertainty quantification significantly outperform Elo. Overall we can safely say that model-based approaches significantly outperform simple methods like Elo and Glicko and are preferable for the considered sports.

Here we have assessed the accuracy of the approaches in predicting match outcomes, as this is a natural task in the context of online decision-making, and can be useful for a variety of purposes including seeding, scheduling and indeed betting. Predictive distributions can also be used for more sophisticated decision-making such as those based on multiple future matches (competition outcomes, promotion/relegation results, etc.).

\subsection{Smoothing (for historical evaluation)}

Smoothing represents an integral subroutine for parameter estimation, though can also be of interest in its own right \citep{Glickman1999, duffield2022online}. In particular, when analysing the historical evolution of a player's skill over time, it is more appropriate to consider the smoothing distributions, rather than the filtering distributions which do not update in light of recent match results.

\begin{figure}[h!]
    \centering
    \begin{minipage}{0.49\linewidth}
        \centering
        \includegraphics[width=\textwidth]{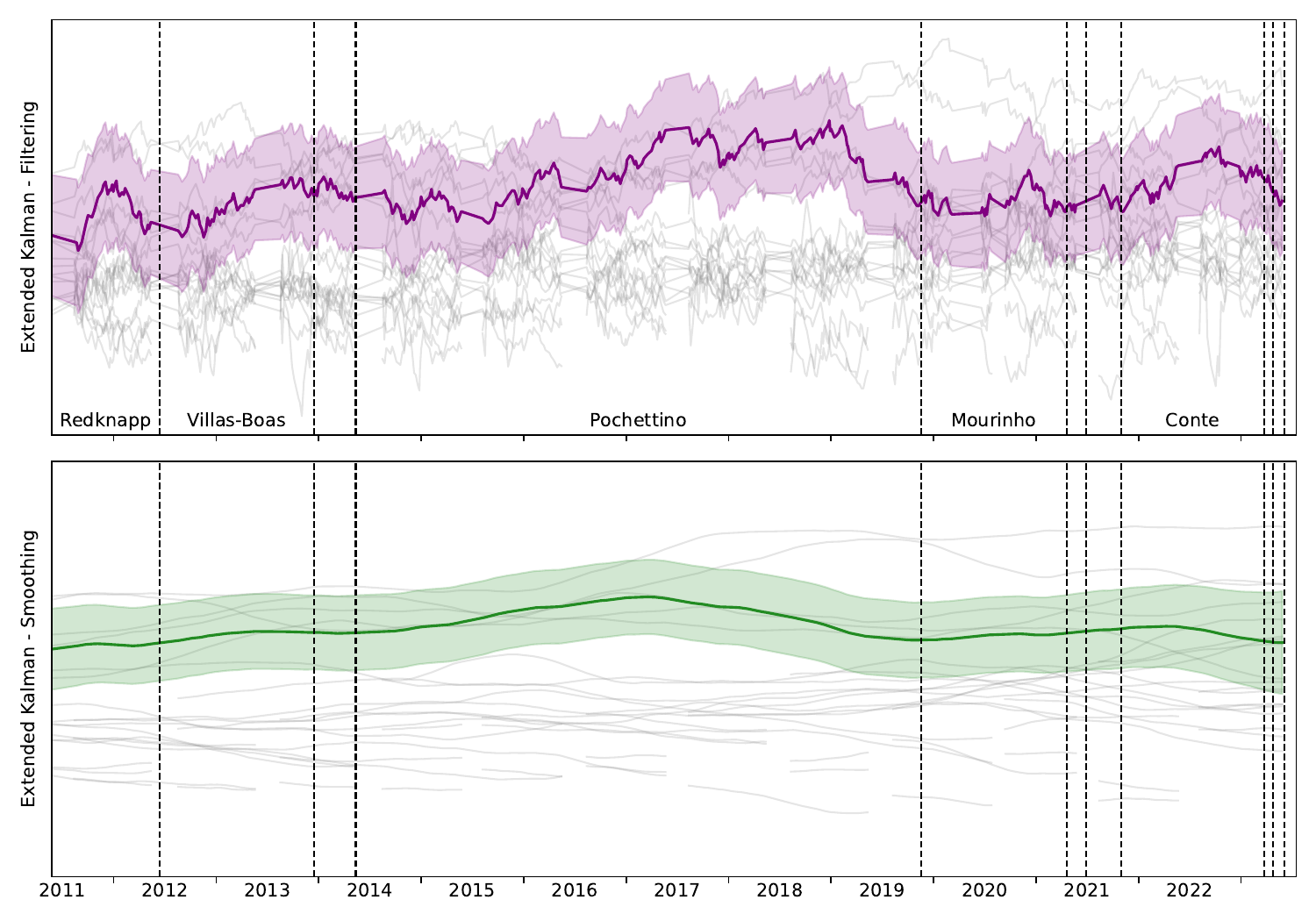}
    \end{minipage}
    \begin{minipage}{0.49\linewidth}
        \centering
        \includegraphics[width=\textwidth]{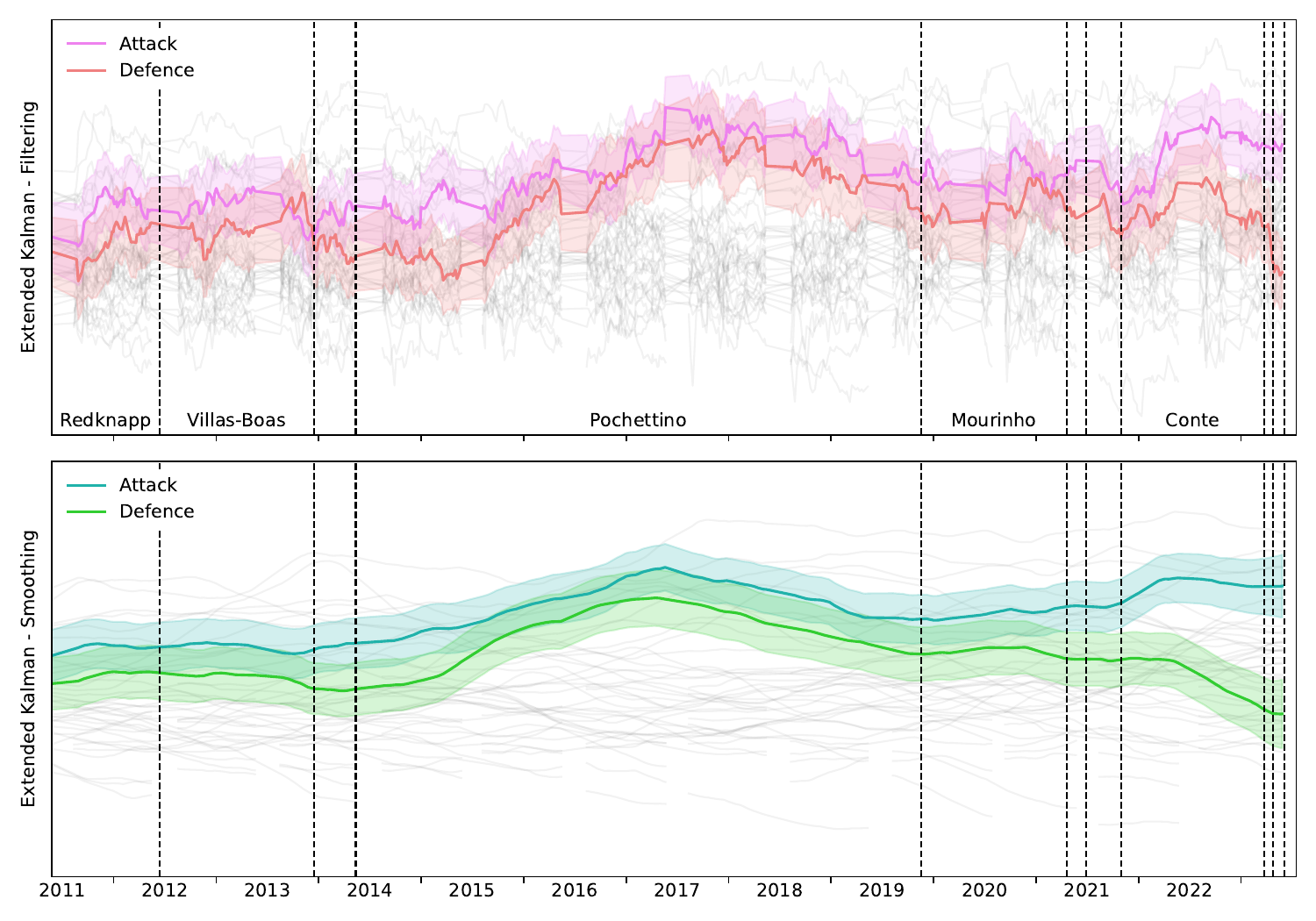}
    \end{minipage}
    \caption{Extended Kalman filtering and smoothing with for Tottenham's EPL matches from 2011-2023. Top: filtering. Bottom: smoothing. Left: sigmoidal likelihood \eqref{specific_SSM}. Right: bivariate Poisson likelihood  \eqref{bivariate_poisson}. Error bars represent one standard deviation with the other teams' mean skills in faded grey. Black dashed lines represent a change in Tottenham manager with long-serving ones named.}
    \label{fig:tottenham}
\end{figure}

On the left of Fig.~\ref{fig:tottenham}, we display the historical evolution of Tottenham's EPL skill rating over time according to \eqref{specific_SSM} and extended Kalman inference. When comparing filtering and smoothing, we immediately see that the smoothing distributions are less reactive, and provide a more realistic trajectory of how a team's underlying skill is expected to evolve over time. Noting that the model permits a certain amount of randomness to occur in each match which can result in surprise results, we observe that the smoothing distributions do a much better job of handling this noise or uncertainty. Fig.~\ref{fig:tottenham} only displays the extended Kalman method, but similar takeaways hold for all of the aforementioned model-based methods (as is highlighted in Fig.~\ref{fig:argentina}).

Historical evaluation of skills can be particularly useful to analyse the impacts of various factors and how they impact the team's underlying skill level, relative to its competitors. In Fig.~\ref{fig:tottenham}, we highlight the different managers or head coaches that have served Tottenham during the time period, depicting their competitors' mean skills in the background. This can be particularly useful in evaluating the ability of the managers and their impact on the team. We observe from the smoothing output that Tottenham's skill rating was in ascendance under Villas-Boas and early Pochettino, before descending towards the end of the Pochettino era (although perhaps not as sharply as the filtering output would suggest). We note that there are likely many further factors influencing the underlying skill that are not highlighted in Fig.~\ref{fig:tottenham}, and also that \eqref{specific_SSM} is relatively simple; it may of course be desirable to build a more complex model which can direct account for such additional factors or data.

In practice, one may want to update full smoothing trajectories in an online fashion so that historical evaluation is possible without having to run full $\bigOh (K)$ backward sweeps. Here, a fixed-lag approximation is a natural option \citep{duffield2022online}.

\subsection{Model Refinement}
Focusing on the case of football, simply modelling match outcomes as draw, home win or away win ignores valuable information recorded in the margin of victory and the number of goals scored.  %We therefore posit we might be able to obtain better predictions through a more refined model that takes this into account and further that we would then also be able to make predictions on matters such as number of goals or home advantage.

As described in \ref{subsec:extensions}, we can intuitively model the number of goals scored in a match via a bivariate Poisson model \citep{karlis2003analysis}, where attack and defence strengths represent the underlying latent player skills. At the bottom of Table~\ref{tab:num_comparison} we can see that our refined model provides better predictions for the match winner than the sigmoidal SSM \eqref{specific_SSM}, despite not being directly optimised for the purpose.

On the right of Fig.~\ref{fig:tottenham}, we visualise the extension of our model for attack and defensive skills, again for both filtering and smoothing. The more expressive model allows us to conclude that the key driver of Tottenham's decrease in overall skill in 2022 was due to a weakening of their defence, whilst their attack remained relatively proficient.

\section{Discussion\label{sec:Discussion}}

In this work, we have advocated for a model-based approach to the skill rating problem. By taking this perspective, we can separate the tasks of modelling and inference. We have detailed several SSMs which are suitable for tackling the problem and are highly interpretable and extensible to suit more sophisticated data pipelines. We have also detailed different approximate inference schemes for analysing such models and discussed their relative strengths and shortcomings. We have conducted thorough case studies on how such methods apply to real data, highlighting a simple workflow, and the different roles of filtering, smoothing, and parameter estimation.

% \subsection*{Advantages over alternative approaches}
%We briefly pause to contrast the state-space model paradigm against alternative approaches to the task of skill rating.

As we have demonstrated, a principled Bayesian approach offers both interpretable and predictive advantages over baseline methods such as Elo and Glicko. These advantages also apply to alternative algorithms that ignore the quality of the opposition, failing to provide predictions for the match outcomes and difficult to extend to more complex data settings or inferential questions. For example, it might be tempting to compare Fig~\ref{fig:tottenham} to a plot of a moving average of Tottenham's points-per-game over the considered period.

%As we have demonstrated, a principled Bayesian approach offers both interpretable and predictive advantages over baseline methods such as Elo and Glicko. These advantages also apply to alternative baseline methods, for example, one might be tempted to compare Fig~\ref{fig:tottenham} to a plot of a moving average of Tottenham's points-per-game over the period. However, such methods ignore the quality of the opposition, fail to provide predictions for the match outcomes and are difficult to extend to more complex data settings or inferential questions.

On the other hand, one might obtain good predictive performance via sophisticated machine learning or black-box methods \citep{menke2008bradley}. However, one would sacrifice the interpretability provided by the state-space model approach, resulting in a model that exhibits difficult-to-challenge assumptions that may fail in subtle ways (such as underfitting or overfitting). The state-space model framework is transparent and through Bayesian marginalisation offers a principled route to uncertainty quantification and a broad range of predictive functionalities (such as the match result via the number of home and away goals in the bivariate Poisson model).

There is of course ample further potential to apply the same procedures to more complex models. Our techniques could even be easily applied point-by-point. For example, one could develop a model for cricket (or baseball) skills where the bowler and batter are the two `players' with an asymmetric likelihood based on $\mathcal{Y} = \left\{\text{wicket}, \text{extras}, 0, 1, 2, 3, 4, 6 \right\}$.

We further remark that point-by-point style state-space models have been considered in the `hot-hand effect' literature \citep{wetzels2016bayesian,otting2020hot,mews2023continuous}. Here, players may enter a `hot' state where they display exceptional performance and significantly affect the match outcome. However, the hot-hand literature primarily focuses on identifying the performance levels of individual players, neglecting cross-player interactions and their combined impact on the match (which may be a suitable assumption for sports such as darts). This focus leads to state-space models that are completely decoupled across players, resulting in a simplified scenario. In contrast, we consider state-space models for which the player skills jointly affect the observations, leading to additional inferential and computational challenges. %Although, it would be possible and indeed interesting to incorporate streak identifying dynamics within our interacting state-space model framework.

% Higher-dimensional representations of player skill also represents a natural extension; e.g. home and away strength, surface-dependent strength for tennis players, cricket batter strength v.s. pace/spin, and more. In this setting, the various quantities for a single player may carry significant correlation, suggesting that the factorial approximation ought be applied \textit{across} player ratings but not \textit{within}. More broadly, this raises considerations about the scalability of the inference techniques with respect to dimension. These considerations also apply to sports which go beyond pairwise observation models (such as those tackled by TrueSkill \citep{Herbrich2006, Minka2018}) although the general joint $\mathsf{Assimilate}$ and $\mathsf{Marginalise}$ framework still applies.

An appealing aspect of our framework is that the user can devote their time to carefully designing their model, and describing their data-generating process in state-space language. Having done this, the general-purpose inference schemes will typically apply directly, enabling the user to easily explore different model and parameter configurations, with the ability to refine the model in an iterative manner \citep{gelman2020bayesian}. The SSM framework also provides the tools (marginal likelihoods) required for model selection in a principled way via Bayesian model selection \citep{wasserman2000bayesian}.

As a result of our specific interests in this problem, we tend to emphasise the role of filtering as a tool for online decision-making, and of smoothing for retrospective evaluation of policies (e.g. assessing coach efficacy, changes in conditions, etc.). For simplicity, we have given a comparatively lightweight treatment of parameter estimation; there are various extensions of our work in this direction which would be worthwhile to examine carefully, e.g. online parameter estimation \citep{cappe2011online,kantas2015particle}, Bayesian approaches \citep{andrieu2010particle}, and beyond.

With regard to practical recommendations, at a coarse resolution, we can offer that i) when speed and scalability are of primary interest, Extended Kalman inference offers a good default, whereas ii) when robustness and flexibility are a greater priority, the fHMM model with graph-based inference has many nice properties. We note that the task of evaluating the relative suitability of \{ Extended Kalman, SMC, HMM, ... \} approaches is not limited to the skill rating setting, and is relevant across many fields and applications wherein state-space models are fundamental. 

%%%%%%%%%%%%%%%%%%%%%%%%%%%%%%%%%%%%%%%%%%%%%%%%%%%%%%%%%%%%%%%%%%%%%%%%%%%%%%%%%%%%%%%%%%%%%%%%%%%%%%%%%%%%%%%%%%%%%%%%%%%%

\section*{Data availability}

All data used is freely available online. Tennis data from \href{http://www.tennis-data.co.uk/alldata.php}{tennis-data.co.uk}. Football data from \href{https://www.football-data.co.uk/englandm.php}{football-data.co.uk} and \href{https://github.com/martj42/international_results}{github.com/martj42/international\_results}. Chess data from \href{https://github.com/huffyhenry/forecasting-candidates}{github.com/huffyhenry/forecasting-candidates}. Reproducible code can be found at \href{https://github.com/SamDuffield/abile}{github.com/SamDuffield/abile}.

%%%%%%%%%%%%%%%%%%%%%%%%%%%%%%%%%%%%%%%%%%%%%%%%%%%%%%%%%%%%%%%%%%%%%%%%%%%%%%%%%%%%%%%%%%%%%%%%%%%%%%%%%%%%%%%%%%%%%%%%%%%%
\section*{Funding}
% \section*{Acknowledgements}
S. Power and L. Rimella were supported by EPSRC grant EP/R018561/1 (Bayes4Health).

%%%%%%%%%%%%%%%%%%%%%%%%%%%%%%%%%%%%%%%%%%%%%%%%%%%%%%%%%%%%%%%%%%%%%%%%%%%%%%%%%%%%%%%%%%%%%%%%%%%%%%%%%%%%%%%%%%%%%%%%%%%%

\bibliographystyle{chicago}
\bibliography{sample.bib}

\appendix

\section{Algorithm Recursions}\label{app:algs}

We here detail the mathematical recursions underlying the methods in the main text. For brevity, we discuss the recursions in the context of the $\bigOh (N \cdot K)$ posterior - (1) in the main text. However, any practical implementation should utilise the memory efficient $\bigOh (N + K)$ match-sparsity posterior - (2) in the main text. Also note that by the factorial approximation, smoothing can be applied independently across players and we therefore omit the player index superscript for the smoothing recursions. A python package with code implementing all of the below can be found at \href{https://github.com/SamDuffield/abile}{github.com/SamDuffield/abile}.

\subsection{Elo-Davidson}\label{app:Elo}

The Elo-Davidson rating system generalises the original Elo \citep{Elo1978} to provide normalised outcome predictions for sports that admit draws \citep{Davidson1970}. The static parameters are a learning rate $K$, a draw propensity parameter $\kappa$ and a scaling parameter $s$ which we set $s=1$.

\subsubsection{Filtering}

\begin{align*}
    x^h_k = x_{k-1}^h + K \cdot \left(\mathbb{I}[y_k = h] + \frac12\mathbb{I}[y_k = \text{draw}] - g\left(\frac{x_{k-1}^h - x_{k-1}^a}{s} ; \kappa\right)
    \right), \\
    x^a_k = x_{k-1}^a + K \cdot \left(\mathbb{I}[y_k = a] + \frac12\mathbb{I}[y_k = \text{draw}] - g\left(\frac{x_{k-1}^a - x_{k-1}^h}{s}; \kappa\right)
    \right),
\end{align*}
where $g(z ; \kappa) = \frac{10^{z} + \kappa / 2}{10^{-z} + 10^{z} + \kappa}$ and the outcome predictions are given by
\begin{align*}
    \mathbf{P}(y_k = h) &\propto 10^\frac{x_{k-1}^h - x_{k-1}^a}{s} \\
    \mathbf{P}(y_k = a) &\propto 10^\frac{x_{k-1}^a - x_{k-1}^h}{s} \\
    \mathbf{P}(y_k = \text{draw}) &\propto \kappa.
\end{align*}

\subsubsection{Smoothing and Parameter Estimation}

The Elo-Davidson recursions are not explicitly model-based and therefore are not accompanied by smoothing recursions or parameter estimation (beyond optimising the predictive performance objective via grid search on $K$ and $\kappa$, as described in the main paper).

\subsection{Glicko}\label{app:Glicko}

The Glicko rating system \citep{Glickman1999} extends Elo to include a time-varying spread variable for each skill rating, i.e. $\gauss(x^i_k\mid \mu^i_k, \sigma^{i \, 2}_k)$.

The static parameters are an initialisation mean $\mu_0$ (which we fix as $\mu_0 = 0$), an initialisation spread $\sigma_0$, a rate parameter $\tau$, a maximum spread parameter $\sigma_\text{max}$ (which we fix to $\sigma_\text{max} = \sigma_0$) and a scaling parameter $s$ (which we fix as $s=1$).

Glicko does not provide normalised outcome predictions for sports with draws; in the next section, we detail how this can be naturally achieved by adopting a state-space model perspective, and conducting approximate inference by means of the Extended Kalman filter \citep{Ingram2021}.

\subsubsection{Filtering}

The Glicko recursions can be thought of as an approximate filtering step (see the main paper for more details on the filtering algorithm)
\begin{align*}
    \text{Filter}_{k-1}^i &= \gauss(x_{k-1}^i \mid \mu_{k-1}^i, \sigma_{k-1}^{i \, 2}), \\
    \text{Predict}^i_k &= \gauss\left(x_{k}^i \mid \mu_{k-1}^i, \min\left(
    \sigma_{k-1}^{i \, 2} + \tau^2 \cdot (t_k - t_{k-1}), \sigma_\text{max}^2
    \right) \right), \\
    \begin{pmatrix}
        \text{Filter}_k^h \\
        \text{Filter}_k^a
    \end{pmatrix}
    &=
    \begin{pmatrix}
        \gauss(x_{k}^i \mid \mu_{k}^h, \sigma_{k}^{h \, 2}) \\
        \gauss(x_{k}^i \mid \mu_{k}^a, \sigma_{k}^{a \, 2})
    \end{pmatrix}
    =
    \mathsf{Assimilate}\left(
    \begin{pmatrix}
        \text{Predict}_k^h \\
        \text{Predict}_k^a
    \end{pmatrix}, y_k
    \right)
\end{align*}
where the exact form of the $\mathsf{Assimilate}$ equations in the last line, as well as normalised outcome predictions (only for sports without draws), can be found in Section 3.3 of \cite{Glickman1999}.

\subsubsection{Smoothing and Parameter Estimation}

In \cite{Glickman1999}, smoothing was conducted by means of the Kalman smoother (which we detail in the next section), and model parameters were specified by numerically minimising the cross-entropy loss between predicted and observed match outcomes.

\subsection{Extended Kalman}\label{app:ExK}

The Extended Kalman approach allows for the inference procedure of Glicko to be generalised to arbitrary state-space models. Note that for generic SSMs which are either non-linear, non-Gaussian, or both, inference by Extended Kalman methods incurs a systematic bias: the true (non-Gaussian) filtering and smoothing distributions are recursively approximated by Gaussian distributions. This bias is separate to the bias which is incurred throughout by the factorial approximation.

To extend Glicko, we can thus re-use the natural state-space model for pairwise comparisons, (5) in the main text and repeated here
\begin{align}\label{specific_SSM}
    m_0 \left( x_0^i \right) = \gauss &\left(x_0^i \mid \mu_0, \sigma_0^2 \right), \quad
    M_{t, t^\prime} \left(x^i_{t}, x^i_{t^\prime} \right) = \gauss \left(x^i_{t^\prime} \mid x^i_{t}, \tau^2 \cdot \left( t^\prime-t \right) \right),
    \\
    G_k \left( y_k \mid x^h, x^a \right)
    &=
    \begin{cases}
        \sigmoid \left(\frac{x^h - x^a + \epsilon}{s}\right) - \sigmoid\left(\frac{x^h - x^a - \epsilon}{s} \right) & \text{if } y_k = \text{draw}, \\
        \sigmoid \left(\frac{x^h - x^a - \epsilon}{s} \right) & \text{if } y_k = \mathrm{h},\\
        1-\sigmoid \left(\frac{x^h - x^a + \epsilon}{s} \right) & \text{if } y_k = \mathrm{a}.
    \end{cases}\nonumber
\end{align}
with static parameters $\theta = (\sigma_0, \tau, \epsilon)$ and fixed $\mu_0=0$ and $s=1$.

\subsubsection{Filtering}

As with Glicko, we have an analytical $\mathsf{Propagate}$ step
\begin{align*}
    \text{Filter}_{k-1}^i &= \gauss(x_{k-1}^i \mid \mu_{k-1}^i, \sigma_{k-1}^{i \, 2}), \\
    \text{Predict}^i_k &= \gauss\left(x_{k}^i \mid \mu_{k-1}^i, 
    \sigma_{k-1}^{i \, 2} + \tau^2 \cdot (t_k - t_{k-1})
    \right),
\end{align*}
noting that we do not synthetically constrain the skill variances as they do.

Due to the non-Gaussian observation model, we cannot conduct the $\mathsf{Assimilate}$ step exactly. Following the Extended Kalman paradigm, we thus proceed by i) computing a second-order Taylor expansion of the log-likelihood $\log G_k$ around the predictive means of $x_k^h$ and $x_k^a$, ii) converting this into a linear-Gaussian approximate likelihood $\hat{G}_k$, i.e.
\begin{align*}
   \log \hat{G}_k(y \mid x^h, x^a) &:= \log G_k(y \mid \mu^h, \mu^a)\\
    &\quad + (x^h - \mu^h, x^a - \mu^a) \cdot \mathrm{g}_k \\
    &\qquad + \frac12 \cdot (x^h - \mu^h, x^a - \mu^a) \cdot \mathrm{H}_k \cdot
    \begin{pmatrix}
        x^h - \mu^h \\ x^a - \mu^a
    \end{pmatrix} &&\in \real,\\
    \mathrm{g}_k &= \nabla_{x^h, x^a} \log G_k(y \mid x^h, x^a)|_{(x^h, x^a) = (\mu^h, \mu^a)} &&\in \real^2, \\
    \mathrm{H}_k &= \nabla_{x^h, x^a}^2 \log G_k(y \mid x^h, x^a)|_{(x^h, x^a) = (\mu^h, \mu^a)} &&\in \real^{2\times2},
\end{align*}
and then finally iii) exactly assimilating this approximate likelihood, i.e.
\begin{align}
    \text{Filter}^{h, a}_k(x^h_k, x^a_k) &\propto \text{Predict}^h_k(x^h_k) \cdot \text{Predict}^a_k(x^a_k) \cdot G_k(y_k \mid x^h_k, x^a_k) \\
    &\approx \text{Predict}^h_k(x^h_k) \cdot \text{Predict}^a_k(x^a_k) \cdot \hat{G}_k(y_k \mid x^h_k, x^a_k), \label{eq:joint_filter}
    \\
    &\propto \gauss \left(
    \begin{pmatrix}
        x^h_k \\
        x_k^a
    \end{pmatrix}
    \mid 
    \begin{pmatrix}
        \mu^h_k \\
        \mu_k^a
    \end{pmatrix}
    ,
    \begin{pmatrix}
        \sigma^{h \, 2}_k & \rho^{h, a}_k \\
        \rho^{h, a}_k & \sigma^{a \, 2}_k
    \end{pmatrix}
    \right), \nonumber
\end{align}
where the final Gaussian distribution is computed by standard linear-Gaussian conjugacy relations; see Chapter 7 in \cite{sarkka2023bayesian}. This is a slightly non-standard version of the extended Kalman filter which linearised the log likelihood directly following \cite{ollivier2018online} as opposed to alternative variants that linearise via Jacobians on forward function to obtain an approximate linear Gaussian state-space model \cite{sarkka2023bayesian} which requires the undesirable additional specification of an observation noise parameter.

Note that it would also be possible to make a first-order Taylor approximation if required for higher-dimensional factorial states (i.e. with $\mathrm{H}_k=0$). The factorial approximation can then be regained through a $\mathsf{Marginalise}$ step which simply consists of reading off $\mu^h_k, \sigma^h_k$ and $\mu^a_k, \sigma^a_k$ and discarding the covariance $\rho^{h, a}_k$.

By using the approximate likelihood $\hat{G}_k$, this approach can provide normalised outcome predictions, i.e.
\begin{align*}
\hat{\mathbf{P}}(y_k \mid y_{1:k-1}) = \int \hat{G}_k(y_k \mid x_k^h, x_k^a) \cdot \text{Predict}^h_k(x_k^h) \cdot \text{Predict}^a_k(x^a_k) \; \mathrm{d}x_k^h \;\mathrm{d}x_k^a.
\end{align*}

\subsubsection{Smoothing}

Given Gaussian filtering approximations and Gaussian dynamics,
\begin{align*}
    M_{k, k+1}(x_k, x_{k+1}) = \gauss(x_{k+1} \mid x_k, \tau^2 \cdot (t_{k+1} - t_k) ),
\end{align*}
smoothing can carried out analytically using the Kalman smoother; see e.g. Section 7.2 in \cite{Chopin2020}. That is, given $\text{Filter}_k(x_k) = \gauss(x_k \mid \mu_k, \sigma_k^2)$ and $\text{Smooth}_{k+1}(x_{k+1}) = \gauss(x_{k+1} \mid \mu_{k+1|K}, \sigma_{k+1|K}^2)$, we have the recursions
\begin{align*}
    \text{Smooth}_{k|K}(x_k) &= \gauss(x_k \mid \mu_{k|K}, \sigma_{k|K}^2), \\
    \mathrm{D}_{k} &= \sigma_k^2 \cdot (\sigma_k^2 + \tau^2 \cdot (t_{k+1} - t_k))^{-1}, \\
    \mu_{k|K} &= \mu_k + \mathrm{D}_{k} \cdot (\mu_{k+1|K} - \mu_k), \\
    \sigma^2_{k|K} &= \sigma^2_k + \mathrm{D}^2_{k} \cdot (\sigma^2_{k+1|K} - \sigma^2_{k} - \tau^2 \cdot (t_{k+1} - t_k) ),
\end{align*}
Additionally, for parameter estimation by expectation-maximisation, it is necessary to compute lag-one autocovariances (i.e. $\mathbf{E}_{\text{Smooth}_{k,k+1|K}}[x_k \cdot x_{k+1}]$); these can be recovered from the same output as
\begin{align*}
    \mu_{k, k+1|K} :=  \mu_{k|K} \cdot \mu_{k+1|K} + \mathrm{D}_{k} \cdot \sigma^2_{k+1|K}.
\end{align*}
The above iterations can then be repeated backwards in time for $k=K-1, \dots, 0$, as well as across the $N$ players (independently, according to the factorial approximation).

\subsubsection{Parameter Estimation}

We can apply an iterative expectation-maximisation algorithm seeking
\begin{align*}
    \argmax_{\theta \in \Theta} \mathbf{P}(y_{1:K} \mid \theta).
\end{align*}
Given a value of the static parameters $\theta_r$, we can run our filtering (storing the filtering statistics) and smoothing algorithms to obtain the smoothing statistics 
\begin{align*}
    \left\{ \mu_{0|K}^{i}, \sigma_{0|K}^{i}, \mu_{0,1|K}^{i}, \dots, \mu_{K-1,K|K}^{i}, \mu_{K|K}^{i}, \sigma_{K|K}^{i} \right\}_{i=1}^N.
\end{align*}
In reality, we only need to obtain $\bigOh (N + K)$ smoothing statistics (i.e. for players at the matchtimes of matches they were involved in), as described in the section of the main paper on match sparsity. For significant ease of notation, we continue here with the $\bigOh (N\cdot K)$ description.

We then look to maximise the surrogate objective - (3) in the main text,
\begin{align}\label{eq:em_q}
    \mathbf{Q} \left( \theta \mid \hat{\theta} \right) := \int \mathbf{P} \left(x_{0:K}^{\left[ N \right]} \mid y_{1:K},\hat{\theta} \right) \cdot \log \mathbf{P}\left( x_{0:K}^{\left[N\right]},  y_{1:K} \mid \theta \right) \, \mathrm{d}x_{0:K}^{\left[N\right]},
\end{align}
Using our recursive Gaussian approximation to the smoothing distribution, we can then extract an approximation to the necessary smoothing statistics, yielding a simple approximation to $\mathbf{Q}(\theta \mid \hat{\theta})$. In particular, in the context of our SSM \eqref{specific_SSM} and Gaussian smoothing statistics, we obtain analytically that
\begin{align*}
    \hat{\sigma}_0^2 &= \frac{1}{N} \sum_{i=1}^N \sigma_{0|K}^{i \, 2} + \mu_{0|K}^{i \, 2}, \\
    \hat{\tau}^2 &= \frac{1}{NK} \sum_{i=1}^N\sum_{k=1}^K \frac{1}{t_k - t_{k-1}}\left[ 
    \sigma_{k-1|K}^{i \, 2} + \mu_{k-1|K}^{i \, 2} - 2 \mu_{k-1,k|K}^{i} + \sigma_{k|K}^{i \, 2} + \mu_{k|K}^{i \, 2}
    \right].
\end{align*}
For parameters corresponding to the observation density and $\epsilon$ parameter, we are required to evaluate integrals of the form
\begin{align*} 
    \int \gauss(z \mid \mu, \sigma^2) \cdot \log \left( \sigmoid(z + \epsilon) - \sigmoid(z - \epsilon)\right) \, dz,
\end{align*}
which cannot be evaluated analytically. However, these can be approximated efficiently using e.g. Gauss-Hermite quadrature \citep{ghq}, and then $\hat{\epsilon}$ can be found using fast univariate numerical optimisation routines.

We then set $\hat{\theta} \leftarrow (\hat{\sigma}_0, \hat{\tau}, \hat{\epsilon})$ and iterate the above filtering, smoothing, and maximisation steps until convergence.

\subsection{TrueSkill2}\label{app:TrueSkill}

The TrueSkill2 approach uses the same SSM \eqref{specific_SSM} as the Extended Kalman approach above, but instead using  the inverse probit sigmoid function $\sigmoid_{\text{IP}}(x) = \Phi(x)$ as (inverse) link function.

\subsubsection{Filtering}

The $\mathsf{Propagate}$ step is again analytically tractable, as in Glicko and the Extended Kalman approach. However, the $\mathsf{Assimilate}$ step is handled differently: instead of Taylor expansion of the log-likelihood, we adopt a moment-matching strategy. In particular, we can use the inverse probit sigmoid and Equation (6) in the main text, repeated here
\begin{align}\label{eq:probit_int}
    \int \gauss(z \mid \mu, \sigma^2) \Phi(z) \, \mathrm{d}z = \Phi \left( \frac{\mu}{\sqrt{1 + \sigma^2}} \right).
\end{align}
This allows us to calculate analytically the marginal means and variances of the non-Gaussian joint filtering distribution \eqref{eq:joint_filter}
\begin{align*}
    \mu_k^h = \mathbf{E}_{\text{Filter}^{h, a}_k}\left[x_k^h\right], \qquad
    \sigma_k^{h \, 2} = \mathbf{Var}_{\text{Filter}^{h, a}_k}\left[x_k^h\right] = \mathbf{E}_{\text{Filter}^{h, a}_k}\left[\left(x_k^h - \mu_k^h\right)^2\right],
\end{align*}
and similarly for $x^a_k$. That is, the above integrals are analytically tractable. As such, we simply represent our approximate filtering distribution by the Gaussian distribution with this mean and variance, still using the factorial approach to decouple approximations across players. This approach is connected to Assumed Density Filtering (ADF) and Expectation Propagation (EP) methods; see later in this supplement for more details.

The identity in \eqref{eq:probit_int} also allows exact computation of the normalised match outcome predictions
\begin{align}\label{eq:match_preds}
\mathbf{P}(y_k \mid y_{1:k-1}) = \int G_k(y_k \mid x_k^h, x_k^a) \cdot \text{Predict}^h_k(x_k^h) \cdot \text{Predict}^a_k(x^a_k) \; \mathrm{d}x_k^h \;\mathrm{d}x_k^a,
\end{align}
for $y_k \in \{\text{draw}, \text{home win}, \text{away win}\}$.

\subsubsection{Smoothing and Parameter Estimation}

These steps can be implemented similarly to the Extended Kalman approach above. Note that in \cite{Minka2018}, they instead use a gradient step (as opposed to analytical maximisation) within the iterative parameter estimation procedure.

\subsection{Sequential Monte Carlo}\label{app:SMC}

Sequential Monte Carlo (SMC) \citep{Chopin2020} takes an importance sampling approach to inference, where a weighted collection of particles is used to approximate otherwise intractable distributions, i.e.
\begin{align*}
    p(x) \approx \sum_{j=1}^J w^j \cdot \delta(x \mid x^j)
\end{align*}
for a normalised collection of weights $\sum_{j=1}^J w^j = 1$. SMC is applicable to general SSMs, although in our context of pairwise comparisons, we again focus on \eqref{specific_SSM}.

\subsubsection{Filtering}
For the $\mathsf{Propagate}$ step, we can assume we have a (weighted) particle approximation to the previous filtering distribution
\begin{align*}
    \text{Filter}^i_{k-1} = \sum_{j=1}^J w^{i \, j}_{k-1} \cdot \delta(x_{k-1}^{i} \mid x_{k-1}^{i \, j}).
\end{align*}
The $\mathsf{Propagate}$ step can then be implemented by simulations, i.e. sampling from the transition kernel $\hat{x}_k^{i \, j} \sim M_{k-1, k}\left(x_{k-1}^{i \, j}, \cdot \right)$ for $j=1,\dots J$ to give
\begin{align*}
    \text{Predict}^i_{k|k-1} = \sum_{j=1}^J w^{i \, j}_{k-1} \cdot \delta( x_k^i \mid \hat{x}_{k}^{i \, j}).
\end{align*}
In certain situations, it can be valuable to instead $\mathsf{Propagate}$ using alternative dynamics, and correct for this discrepancy using importance weighting; we do not pursue this here, but see \cite{doucet2000sequential} for details.

The $\mathsf{Assimilate}$ step is implemented by i) arbitrarily pairing together particles for the two competing players ii) adjusting the importance weights based on the observation likelihood, iii) normalising and then iv) (optionally) conducting a resampling step. The reweighting (and normalising) gives an approximation to the joint filtering distribution
\begin{align*}
    \text{Filter}^{h,a}_k &= \sum_{j=1}^J \hat{w}_k^j \cdot \delta\left(x^{h}_k, x^{a}_k  \mid \hat{x}^{h \, j}_k, \hat{x}^{a \, j}_k\right),
    \\
    \hat{w}_k^j &\propto w^{h \, j}_{k-1} \cdot w^{a \, j}_{k-1} \cdot G_k \left(y_k \mid \hat{x}^{h \, j}_k, \hat{x}^{a \, j}_k\right).
\end{align*}
This can then be followed by a resampling step, giving
\begin{align*}
    \text{Filter}^{h,a}_k &\approx \sum_{j=1}^J w_k^j \cdot \delta\left(x^{h}_k, x^{a}_k  \mid x^{h \, j}_k, x^{a \, j}_k\right),
\end{align*}
where $w_k^j \equiv \frac{1}{J}$ and $x^{h}_k, x^{a}_k \sim \text{Filter}^{h,a}_k$ if resampling is applied; otherwise $w_k^j = \hat{w}_k^j$ and $x^{h}_k, x^{a}_k = \hat{x}^{h}_k, \hat{x}^{a}_k$. The resampling operation ensures that low-probability particles are not carried through to the next iteration and stabilises the weights, thereby ensuring a suitably diverse particle approximation and good long-time stability properties. The simplest resampling approach is to apply multinomial resampling at every $\mathsf{Assimilate}$ step (as used in our experiments); more sophisticated resampling methods including stratified resampling and adaptive approaches are also common in practice \citep{douc2005comparison}.

The $\mathsf{Marginalise}$ step can then be applied be simply unpairing the joint $\text{Filter}^{h,a}_k$ in line with the factorial approximation, i.e.
\begin{align*}
    \text{Filter}^{h}_k = \sum_{j=1}^J w_k^j \cdot \delta\left(x^{h}_k \mid x^{h \, j}_k\right), \qquad
    \text{Filter}^{a}_k = \sum_{j=1}^J w_k^j \cdot \delta\left(x^{a}_k \mid x^{a \, j}_k\right).
\end{align*}

Note that the $\mathsf{Assimilate}$ step could be made more robust by considering all possible $J^2$ pairings (as opposed to the arbitrary $J$ pairings of skills) between the two players. 

Normalised match outcome predictions can be obtained by direct particle approximation to \eqref{eq:match_preds}.

\subsubsection{Smoothing}

Smoothing can be applied by use of \textit{backward simulation} \citep{godsill2004monte}, which re-uses the filtering particle approximations to provide an approximation to full smoothing trajectories $\text{Smooth}_{0:K|K} \approx \frac1J \sum_{j=1}^J \delta \left(x_{0:K} \mid x^j_{0:K|K} \right)$. Assuming that we have an unweighted particle approximation to $\text{Smooth}_{k+1|K} \approx \frac1J \sum_{j=1}^J \delta \left(x_{k+1} \mid x^j_{k+1|K} \right)$ and a weighted particle approximation to $\text{Filter}_{k} \approx \sum_{j=1}^J  w_k^j \cdot \delta\left(x_{k} \mid x^j_{k} \right)$, we can then sample to form an approximation to $\text{Smooth}_{k, k+1|K}$
\begin{align*}
    x_{k|K}^j &\sim \sum_{j^\prime=1}^J  w_{k}^{j^\prime \leftarrow j} \cdot \delta\left(x_{k} \mid x^{j^\prime}_{k} \right), \\
    \text{where} \quad w_{k}^{j^\prime \leftarrow j} &\propto w_{k}^{j^\prime} \cdot M_{k, k+1} \left(x^{j^\prime}_{k}, x^{j}_{k+1|K}\right).
\end{align*}
This procedure can then be iterated for $k=K-1, \dots, 0$ to provide an approximation to $\text{Smooth}_{0:K|K}$. The standard backward simulation detailed above has complexity $\bigOh (J^2)$; note however that methods based on rejection sampling \citep{douc2011sequential} and MCMC \citep{dau2023backward} have been developed in order to reduce the complexity to $\bigOh (J)$.

\subsubsection{Parameter Estimation}

For a particle approximation to $\text{Smooth}_{0:K|K} \approx \frac1J \sum_{j=1}^J \delta \left(x_{0:K} \mid x^j_{0:K|K} \right)$, maximisation of the surrogate objective \eqref{eq:em_q} is tractable for $\sigma_0$ and $\tau$ (as in the Gaussian approximation case). This yields the updates
\begin{align*}
    \hat{\sigma}_0^2 &= \frac{1}{NJ} \sum_{i=1}^N \sum_{j=1}^J x_{0|K}^{i \, j \, 2}, \\
    \hat{\tau}^2 &= \frac{1}{NTJ}  \sum_{i=1}^N \sum_{k=1}^K \sum_{j=1}^J \frac{1}{t_k - t_{k-1}} \left(
    x^{i \, j}_{k|K} - x^{i \, j}_{k-1|K}
    \right)^2.
\end{align*}
Similarly to before, we update $\epsilon$ by univariate numerical maximisation methods.

\subsection{Factorial Hidden Markov Model}\label{app:fHMM}

In an HMM approach, the player skill ratings take values in a discrete set $\mathcal{X} = \left[ S \right]$ for some $S \in \mathbf{N}$, which leads to a closed form for the operations $\left\{ \mathsf{Propagate}, \mathsf{Assimilate}, \mathsf{Bridge} \right\}$. However, as mentioned in the main text, the cost of these operations might be exponentially large in the dimension of the state space, and so computationally unfeasible. We can then adopt the decoupling approximation and exploit the factorial structure in the dynamics of the fHMM \citep{rimella2022exploiting}. For the decoupling approximation, the only thing we have to define is the form of $\left\{ \mathsf{Propagate}, \mathsf{Assimilate}, \mathsf{Bridge}, \mathsf{Marginalise}\right\}$, which will then lead to the recursive algorithm. Note also that the flexibility of the fHMM means that we are not limited to specific choices of the transition kernel $M_{k-1,k}$ and emission distribution $G_k$, and so we keep them general in the following sections. That is, we assume only that $M_{k-1,k}$ is an $S \times S$ stochastic matrix and that for any $y \in \mathcal{Y}$, $G_k(y \mid \cdot, \cdot)$ is an $S \times S$ matrix.

\subsubsection{Assembling the kernel}

Before starting, it is important to briefly explain why explicitly assembling the transition kernel $M_{t,t^\prime}$ is inadvisable unless strictly necessary. Recall that we define the dynamics in terms of the generator matrix $Q_S$, and as such $M_{t, t^\prime}$ is expressible as matrix exponential, i.e.
\begin{align*}
    M_{t,t^\prime} = \exp \left( \tau_d \cdot (t^\prime -t) \cdot Q_S \right).
\end{align*}
In principle, computing this matrix for each $t, t^\prime$ incurs a cost of $\bigOh(S^3)$. However, this is wasteful: for filtering purposes, we only need to access $M_{t, t^\prime}$ through its action on probability vectors, which costs only $\bigOh (S^2)$ once the matrix is available. We will thus try to describe these matrices in a way which enables these rapid computations.

In particular, when possible, it is worthwhile to instead pre-compute a diagonalisation of $Q_S$, i.e. identify an invertible matrix $\Psi_S$ and diagonal matrix $\Lambda_S$ such that
\begin{align*}
    Q_S = \Psi_S^{-1} \cdot \Lambda_S \cdot \Psi_S,
\end{align*}
noting that for self-adjoint $Q_S$, $\Psi_S$ will be orthogonal. For the example considered in the main text, we detail this decomposition in Appendix \ref{app:CTRW}.

Given such a decomposition, the transition kernel then takes the form
\begin{align*}
    M_{t, t^\prime} &= \Psi_S^{-1} \cdot \exp \left( -\tau_d \cdot (t^\prime-t) \cdot \Lambda_S \right) \cdot \Psi_S.
\end{align*}
We can thus compute expressions like $\pi M_{t, t^\prime}$ by
\begin{align}
    \pi_0 &= \pi \\
    \pi_1 &= \pi_0 \cdot \Psi_S^{-1} \\
    \pi_2 &= \pi_1 \cdot \exp \left( -\tau_d \cdot (t^\prime-t) \cdot \Lambda_S \right) \\
    \pi_3 &= \pi_2 \cdot \Psi_S.
\end{align}
The first and third steps are standard matrix-vector products, costing $\bigOh (S^2)$, whereas the second step is a diagonal matrix-vector product, costing only $\bigOh (S)$. We thus see that a judicious one-off matrix decomposition costing $\bigOh(S^3)$ allows us to reduce the cost of all future filtering updates to $\bigOh(S^2)$.

Note that in this framework assembling $M_{t, t^\prime}$ has a computational cost of $\bigOh (S^3)$, while vector matrix multiplication with $M_{t, t^\prime}$ remains $\bigOh (S^2)$.

In the next sections, we will see that filtering is not an issue as it requires vector-matrix multiplication with the kernel, while vanilla smoothing requires elementwise products with the transition kernel and so does not simplify in a straightforward vector-matrix operation. However, we can modify the smoothing recursions to require vector-matrix operations only.

\subsubsection{Filtering}
Here we explain the general step $k$ in the filtering recursion in the context of the $\bigOh(N \cdot K)$ representation. This requires performing a $\mathsf{Propagate}$ step and $\mathsf{Assimilate}$ on players $h,a$.

The $\mathsf{Propagate}$ step at time $k$ consists of propagating the filtering distribution through the transition kernel. In finite state spaces, this is a straightforward matrix-vector operation. In particular, for $i\in [N]$, we do
\begin{align*}
    \mathrm{Predict}_{k \mid k-1}^{i} &= \mathsf{Propagate}\left(\mathrm{Filter}_{k-1}^{i}; M_{{k-1}, k}\right) \\
    &= \mathrm{Filter}_{k-1}^{i} \cdot M_{{k-1}, k},
\end{align*}
noting that we represent probability vectors as row vectors.

$\mathsf{Assimilate}$ requires a joint update on both players $h$ and $a$, and so outputs the joint distribution $\mathrm{Filter}^{h, a}_{k}$ in the form of a ${S\times S}$ matrix with elements summing to $1$. Precisely, given the current match outcome $y_k$, the joint filter takes the form
\begin{align*}
     \mathrm{Filter}^{h, a}_{k} &= \mathsf{Assimilate} \left( \left(\begin{array}{c}
    \mathrm{Predict}^{h}_{k|k-1}\\
    \mathrm{Predict}^{a}_{k|k-1}
    \end{array}\right); G_{k} \right) \\
    &\propto \left ( \left ( \mathrm{Predict}_{k \mid k-1}^{h}\right )^{\mathrm{T}}  \cdot \mathrm{Predict}_{k \mid k-1}^{a} \right ) \odot G_k(y_k \mid \cdot,\cdot),   
\end{align*}
where the operation $\odot$ denotes elementwise product between matrices. Observe that on the left hand side of $\odot$ we have the joint prediction on the skill level of $h$ and $a$, while on the right hand side we have the emission score from the match result. It is also important to notice that we have a proportional sign, meaning that we need to normalize our computation to a joint distribution on ${S\times S}$.

Noting that our $\mathsf{Propagate}$ operator acts on player-wise filtering distribution, it remains to convert $\mathrm{Filter}^{h, a}_{k}$ into a product of marginals. To do so, we simply compute the marginals along $h$ and $a$, i.e. sum over the columns and sum over the rows
\begin{align*}
\left(\begin{array}{c}
    \mathrm{Filter}^{h}_{k}\\
    \mathrm{Filter}^{a}_{k}
    \end{array}\right) 
    = \left(\begin{array}{c}
    \mathsf{Marginalise} \left ( \mathrm{Filter}^{h, a}_{k}; h \right )\\
    \mathsf{Marginalise} \left ( \mathrm{Filter}^{h, a}_{k}; a \right )
    \end{array}\right)
    = \left(\begin{array}{c}
     \left( \mathrm{Filter}^{h, a}_{k} \cdot 1_S^\top \right)^\top\\
    1_S \cdot \mathrm{Filter}^{h, a}_{k}
    \end{array}\right),
\end{align*}
where $1_S$ is the $S$-dimensional row vector of ones.

Overall, the computational cost of a filtering step is $\bigOh (S^2)$ per each player, as it requires linear algebra operation on $S$-dimensional vectors and $S \times S$-dimensional matrices. Note that the we are omitting the $\bigOh(S^3)$ cost of the initial matrix decomposition (as it has a complexity independent of both $N$ and $K$), as well as the $\bigOh(S^3)$ cost of assembling the matrices $M_{t, t^\prime}$, which (as described above) is not necessary for either filtering or prediction.

Normalised match outcome predictions can be obtained by evaluating the integral \eqref{eq:match_preds} exactly via linear algebra.

\subsubsection{Smoothing}\label{sec:smoothing_discrete}

This section details a general step of the smoothing algorithm. Recall that by our decoupling approximation, smoothing can carried out independently across players. As such, we temporarily drop the player indices from our notation.

To run the $k$th backward step of the smoothing algorithm, we require the row vectors $\mathrm{Filter}_{k}$ and $\mathrm{Smooth}_{k+1 \mid K}$, representing the filtering distribution of the skills of a player at time $k$, and the smoothing distribution of the skills of the same player at time $k+1$. For a general smoothing step we need the operations $\left \{ \mathsf{Bridge}, \mathsf{Marginalise} \right \}$, where $\mathsf{Bridge}$ generates a joint distribution on the skills of the players on time steps $k,k+1$ and $\mathsf{Marginalise}$ converts this joint distribution in a marginal over earlier time step $k$. As for the filtering step, all operations rely only on basic linear algebra routines:
\begin{align*}
    \begin{split}
    \mathrm{Smooth}_{k,k+1 \mid K} &= \mathsf{Bridge} \left( \mathrm{Filter}_{k}, \mathrm{Smooth}_{k+1 \mid K}; M_{k,{k+1}}\right) \\
    &=  \left[ \mathrm{Filter}_{k}^\top \cdot \left ( \mathrm{Smooth}_{k+1 \mid K} \fatslash \mathrm{Predict}_{k+1 \mid k} \right ) \right]  \odot  M_{k, k+1} \\
    \mathrm{Smooth}_{k \mid K} &= \mathsf{Marginalise} \left( \mathrm{Smooth}_{k,k+1 \mid K}; k \right) = \left( \mathrm{Smooth}_{k, k+1 \mid K} \cdot 1_S^\top \right)^\top,
    \end{split}
\end{align*}
where $\fatslash$ denotes elementwise division.

Given access to the transition matrix $M_{k, k + 1}$, one sees that the smoothing step has a computational cost of $\bigOh(S^2)$; it thus remains to assemble this matrix, at a cost of $\bigOh(S^3)$.

However, there is an alternative formulation of the smoothing step which avoids assembling the joint distribution by merging the $\mathsf{Marginalise}$ and $\mathsf{Bridge}$ steps:
\begin{align*}
\begin{split}
    \mathrm{Predict}_{k+1 \mid k} &= \mathrm{Filter}_{k} \cdot M_{k, k+1} ,\\
    \mathrm{Smooth}_{k \mid K} &= \mathrm{Filter}_{k} \odot \left [ \left ( \mathrm{Smooth}_{k+1 \mid K} \fatslash \mathrm{Predict}_{k+1 \mid k} \right ) \cdot M_{k, k+1}^\top \right ].
\end{split}
\end{align*}
The derivation comes from carefully reordering the operations in such a way that we perform the marginalisation on $k+1$ first, indeed
\begin{equation}
    \begin{split}
       &\left[ \mathrm{Filter}_{k}^\top \cdot \left ( \mathrm{Smooth}_{k+1 \mid K} \fatslash \mathrm{Predict}_{k+1 \mid k} \right ) \right]  \odot  M_{k, k+1} \\
       &= \left ( \mathrm{Filter}_{k}^\top \cdot  1_S \right )  \odot  M_{k, k+1} \odot \left [ 1_S^\top \cdot \left ( \mathrm{Smooth}_{k+1 \mid K} \fatslash \mathrm{Predict}_{k+1 \mid k} \right ) \right ] 
    \end{split}
\end{equation}
and 
\begin{equation}
    \begin{split}
       & \left \{ \left ( \mathrm{Filter}_{k}^\top \cdot  1_S \right )  \odot  M_{k, k+1} \odot \left [ 1_S^\top \cdot \left ( \mathrm{Smooth}_{k+1 \mid K} \fatslash \mathrm{Predict}_{k+1 \mid k} \right ) \right ] \right \} \cdot 1_S^\top\\
       &= \mathrm{Filter}_{k}^\top \odot \left( \left \{ M_{k, k+1} \odot \left [ 1_S^\top \cdot \left ( \mathrm{Smooth}_{k+1 \mid K} \fatslash \mathrm{Predict}_{k+1 \mid k} \right ) \right ]  \right \} \cdot 1_S^\top \right)  \\
       &= \mathrm{Filter}_{k}^\top \odot \left [ M_{k, k+1} \cdot \left ( \mathrm{Smooth}_{k+1 \mid K} \fatslash \mathrm{Predict}_{k+1 \mid k} \right )^\top \right ] 
    \end{split}
\end{equation}
where the first equality follows from noting that $\left ( \mathrm{Filter}_{k}^\top \cdot  1_S \right )$ is constant within rows, and the second from noting that $1_S ^\top\cdot \left ( \mathrm{Smooth}_{k+1 \mid K} \fatslash \mathrm{Predict}_{k+1 \mid k} \right )$ is constant within columns. This alternative perspective on the smoothing step is crucial, as by avoiding the assembly of the transition kernel, we are able to control the cost of smoothing iterations at the much lower $\bigOh (S^2)$.

\subsubsection{Parameter Estimation}

In this section we make explicit the dependence of the model on the parameters $\theta$. In our setting, the parameters are also decoupled, with respect to their impact initial distribution, transition kernel and emission distribution. That is, with $\theta = ( \sigma_d, \tau_d, \epsilon_d )$, only $\sigma_d$ influences the initial distribution, only $\tau_d$ the transition kernel, and only $\epsilon_d$ the emission distribution, with .

Towards implementing the EM algorithm, we recall the definition of $\mathbf{Q}(\theta|\hat{\theta})$ and simply substitute the true smoothing distribution with our decoupled approximation. The resulting $\mathbf{Q}$ function which we seek to maximise may then be expressed as
\begin{align*}
    \mathbf{Q}(\theta|\hat{\theta}) &= \int \mathbf{P} \left( x_{0:K}^{[N]} \mid {y}_{1:K}, \hat{\theta}\right) \cdot \log \mathbf{P} \left( x_{0:K}^{[N]} , {y}_{1:K} \mid {\theta} \right) \, \mathrm{d} x_{0:K}^{[N]} \\
    &= \mathbf{Q}_1 ( \sigma_d | \hat{\theta}) + \mathbf{Q}_2 ( \tau_d | \hat{\theta}) + \mathbf{Q}_3 ( \epsilon_d | \hat{\theta}),
\end{align*}
where
\begin{align*}
    \mathbf{Q}_1 ( \sigma_d | \hat{\theta}) &= \sum_{i \in [N]} \int \mathbf{P} \left( x_0^i \mid {y}_{1:K}, \hat{\theta}\right) \cdot \log m_{0}  \left( {x}_{0}^{i} \mid \sigma_d \right) \, \mathrm{d} x_0^i \\
    \mathbf{Q}_2 ( \tau_d | \hat{\theta}) &= \sum_{i \in [N]}  \sum_{k=1}^K \int \mathbf{P} \left( x_{k-1}^i, x_k^i \mid {y}_{1:K}, \hat{\theta}\right) \cdot \log M_{k-1, k} \left ( x^i_{k-1}, x^i_{k} \mid \tau_d \right ) \, \mathrm{d} x^i_{k-1} \, \mathrm{d} x^i_{k} \\
    \mathbf{Q}_3 ( \epsilon_d | \hat{\theta}) &= \sum_{k=1}^K \int \mathbf{P} \left( x_k^{h(k)}, x_k^{a(k)} \mid {y}_{1:K}, \hat{\theta}\right) \cdot \log G_k \left ( y_k \mid x_k^{h(k)}, x_k^{a(k)} \mid \epsilon_d \right ) \, \mathrm{d} {x}_{k}^{h(k)} \mathrm{d} {x}_{k}^{a(k)}.
\end{align*}
We can then minimise each of these objective functions separately.

One practical difficulty is that in order to construct $\mathbf{Q}_2$, we require access to the two-step joint smoothing distributions, i.e. $\mathrm{Smooth}_{k - 1, k \mid K}$. Assembling this distribution incurs an undesirable cost of $\bigOh (S^3)$, which we prefer to avoid. As such, we proceed with the following hybrid method:
\begin{enumerate}
    \item Compute the single-step smoothing laws using the $\bigOh (S^2)$ algorithm. 
    \item Use the single-step smoothing laws to update the parameters of the initial distribution and emission distribution in closed form.
    \item Compute the gradient of $\mathbf{Q}_2 (\tau_d \mid \hat{\theta})$ with respect to $\tau_d$, and
    \item Update $\tau_d$ by applying a gradient step.
\end{enumerate}

We now detail how this gradient is computed. Focusing on $\mathbf{Q}_2$, we expand the joint smoothing distribution to write
\begin{align*}
    \begin{split}
        \frac{\partial\mathbf{Q}_2 \left( \tau_d | \hat{\theta}\right)}{\partial\tau_d} & = \sum_{i\in\left[N\right]} \sum_{k=1}^{K} \sum_{a,b\in\left[S\right]} \frac{\frac{\partial}{\partial\tau_d} M_{k-1,k} \left(a,b\mid\tau_d\right)}{M_{k-1,k}\left(a,b\mid\tau_d\right)} \cdot \mathrm{Smooth}_{k-1, k \mid K}^{i} \left(a,b \mid \hat{\theta} \right)\\
        & = \sum_{i\in\left[N\right]} \sum_{k=1}^{K} \sum_{a,b\in\left[S\right]}\frac{\frac{\partial}{\partial\tau_d}M_{k-1,k} \left( a, b \mid \tau_d \right)}{M_{k - 1, k} \left( a, b \mid \tau_d\right)}\\
        & \cdot \frac{\mathrm{Filter}_{k-1}^{i}\left(a\mid\hat{\theta}\right) \cdot \mathrm{Smooth}_{k\mid K}^{i} \left(b\mid\hat{\theta}\right) \cdot M_{k-1,k}\left( a, b \mid \hat{\tau}_d \right)}{\sum_{\bar{a},\bar{b}\in\left[S\right]} \mathrm{Filter}_{k-1}^{i}\left(\bar{a}\mid \hat{\theta} \right) \cdot \mathrm{Smooth}_{k\mid K}^{i} \left(\bar{b}\mid \hat{\theta} \right) \cdot M_{k-1,k} \left(\bar{a},\bar{b} \mid \hat{\tau}_d \right)}.
    \end{split}
\end{align*}
This expression can be simplified further if we consider its value at $\tau_d = \hat{\tau}_d$. We obtain
\begin{equation}
    \begin{split}
        &\frac{\partial \mathbf{Q}_2 \left( \tau_d |\hat{\theta}\right)}{\partial\tau_d} = \sum_{i \in \left[ N \right]} \sum_{k=1}^{K} \sum_{a, b \in \left[ S \right]} 
        \left[\frac{\partial}{\partial\tau_d} M_{k-1,k} \left(a,b \mid \hat{\tau}_d \right)\right] \\
        &\cdot
        \frac{\mathrm{Filter}_{k-1}^{i}\left(a\mid\hat{\theta}\right) \cdot \mathrm{Smooth}_{k\mid K}^{i}\left(b \mid \hat{\theta}\right)}{\sum_{\bar{a},\bar{b}\in\left[S\right]} \mathrm{Filter}_{k-1}^{i}\left(\bar{a}\mid\hat{\theta}\right) \cdot \mathrm{Smooth}_{k\mid K}^{i}\left(\bar{b}\mid\hat{\theta}\right)\cdot M_{k-1,k} \left(\bar{a},\bar{b}\mid \hat{\tau}_d \right)}.    
    \end{split}
\end{equation}
Now, suppose that $Q_{S} = \Psi_{S}^{-1} \cdot \Lambda_{S} \cdot \Psi_{S}$, and that $\tau_d = \hat{\tau}_d$ is a rate parameter, i.e.
\begin{align*}
    M_{t,t^{\prime}}\left(\cdot\mid\tau_d\right)=\exp\left(\tau_{d}\cdot\left(t^{\prime}-t\right)\cdot Q_{S}\right).
\end{align*}
It then holds that
\begin{align*}
    \frac{\partial}{\partial\tau_d}M_{k-1,k} \left(a,b \mid \hat{\tau}_d \right)	&=\frac{\partial}{\partial\tau_{d}} \exp \left(\tau_{d} \cdot \left(t_{k}-t_{k-1}\right) \cdot Q_{S}\right) \\
	&= \exp \left(\tau_{d}\cdot\left(t_{k}-t_{k-1}\right)\cdot Q_{S}\right)\cdot\left(t_{k}-t_{k-1}\right)\cdot Q_{S} \\
	&= \Psi_{S}^{-1}\cdot\left[\left(t_{k}-t_{k-1}\right)\cdot\Lambda_{S}\cdot\exp\left(\tau_{d}\cdot\left(t_{k}-t_{k-1}\right)\Lambda_{S}\right)\right]\cdot\Psi_{S},
\end{align*}
where the term in brackets is again a diagonal matrix. The gradient can then be assembled as follows
\begin{align*}
    \mathrm{F}_{k-1}^{i}	&\leftarrow \left( \mathrm{Filter}_{k-1}^{i} \right) \cdot \Psi_{S}^{-1} \\
    \mathrm{S}_{k}^{i}	&\leftarrow \left( \mathrm{Smooth}_{k\mid K}^{i} \right) \cdot \Psi_{S}^{-1} \\
    \tilde{\Lambda}_{S}^{k}	&\leftarrow \left( t_{k}-t_{k-1}\right) \cdot \Lambda_{S} \cdot \exp \left( \tau_{d} \cdot \left( t_{k}-t_{k-1} \right) \Lambda_{S}\right) \\
    \mathrm{N}_{k-1,k}^{i}	&= \mathrm{F}_{k-1}^{i} \cdot \tilde{\Lambda}_{S}^{k} \cdot \left( \mathrm{S}_{k}^{i} \right)^{\top} \\
    \mathrm{D}_{k-1,k}^{i}	&= \mathrm{F}_{k-1}^{i} \cdot \Lambda_{S} \cdot \left( \mathrm{S}_{k}^{i} \right)^{\top} \\
    \frac{\partial \mathbf{Q}_2 \left( \tau_d | \hat{\theta}\right)}{\partial\tau_d} &= \sum_{i\in\left[N\right]} \sum_{k=1}^{K} \frac{\mathrm{N}_{k-1,k}^{i}}{\mathrm{D}_{k-1,k}^{i}}, \\
\end{align*}
wherein the first two terms can be constructed in time $\bigOh (S^2)$, and the subsequent three terms in time $\bigOh (S)$. For generator matrices with more elaborate dependence on parameters, we expect that more specialised routines may be necessary. This is largely a by-product of the non-commutativity of matrices, which complicates the computation of expressions like $\nabla_{\theta}\exp \left( Q\left(\theta\right)\right)$.

\section{Continuous time Random Walk for Finite State Spaces}\label{app:CTRW}

For the fHMM model used in Section 4.5 of the main text, we take our dynamics to be the Continuous-Time Reflected Random Walk on $\left[ S \right]$, which we construct as follows. Let $P_{S}$ denote the transition matrix for the Discrete-Time Reflected Random Walk on $\left[ S \right]$, i.e. 
\begin{align*}
    \left(P_{S}\right)_{i,j} = 
    \begin{cases}
        \frac{1}{2} & 1<i,j<S,\left|i-j\right| = 1 \\
        \frac{1}{2} & \left(i,j\right)\in\left\{ \left(1,2\right),\left(S,S-1\right)\right\} \\
        0 & \text{otherwise},
    \end{cases}
\end{align*}
noting that this process admits the normalised uniform measure on $\left[ S \right]$ as its unique invariant probability measure, and is even reversible with respect to it. As such, we know that $P_{S}$ admits an eigenvalue decomposition of the form $P_{S} = \Psi_{S}^\top \cdot \tilde{\Lambda}_{S} \cdot \Psi_{S}$, where $\Psi_{S}$ is orthogonal, and $\tilde{\Lambda}_{S}$ is diagonal. In fact, $\left(\Psi_{S}, \tilde{\Lambda}_{S}\right)$ can be computed analytically as
\begin{align*}
    \left(\tilde{\Lambda}_{S}\right)_{i,i}	&= \cos\left(\frac{2\cdot\pi\cdot\left(i-1\right)}{S}\right) \\
    \left(\Psi_{S}\right)_{i,j}	&=
    \begin{cases}
        S^{-1/2} & i=1\\
        \left(S/2\right)^{-1/2} \cdot \cos\left(\frac{\pi\cdot\left(i-1\right) \cdot \left(2\cdot j-1\right)}{S}\right) & 1<i\leqslant S.
    \end{cases}
\end{align*}
We now define the continuous-time process as the Poissonisation of the discrete-time process with transition matrix $P_S$, whose generator matrix is obtained as $Q_{S} = P_{S} - I_{S}$. One then sees that $Q_{S}$ admits the eigenvalue decomposition $Q_{S} = \Psi_{S}^\top \cdot \Lambda_{S} \cdot \Psi_{S}$, with $\Lambda_{S}=\tilde{\Lambda}_{S}-I_{S}$, which is again fully explicit.

\section{Expectation-Propagation and Assumed Density Filtering}\label{app:EP}

Assumed Density Filtering (ADF) is a general tool for approximate filtering; see Chapter 1 of \cite{minka2001family} for some overview and many references. In a particularly narrow form, ADF applies to state-space models for which i) there exists a parametric family $\mathcal{F}$ of distributions which is preserved by the dynamics, but ii) the observation model is not necessarily conjugate to this family. Approximating the filtering and predictive distributions by elements of $\mathcal{F}$, ADF then proceeds by implementing the $\mathsf{Propagate}$ step exactly, and approximating the $\mathsf{Assimilate}$ step. When the parametric family $\mathcal{F}$ is an exponential family of distributions, this approximation is typically implemented by a moment-matching step based on the sufficient statistics of the exponential family, i.e.
\begin{align*}
    \mathrm{ExactFilter} &= \mathsf{ExactAssimilate} \left( \mathrm{Predict}, G \right) \\
    \mathrm{Filter} &= \mathsf{MomentMatch}_\mathcal{F} \left( \mathrm{ExactFilter} \right),
\end{align*}
so that one might write $\mathsf{Assimilate} = \mathsf{MomentMatch}_\mathcal{F} \circ \mathsf{ExactAssimilate}$.

A broader interpretation of ADF would allow for non-conjugate dynamics which are then projected back onto the parametric family, implemented by e.g. $\mathsf{Propagate} = \mathsf{MomentMatch}_\mathcal{F} \circ \mathsf{ExactPropagate}$. In general, the feasibility of ADF for any specific model is constrained only by the availability of a suitable parametric family of distributions, and the user's ability to compute the relevant measure projections (by e.g. computing moments).

Expectation Propagation (EP) is an extension of ADF to graphical models which do not necessarily admit a temporal (or even tree-based) structure; see e.g. \cite{minka2013expectation}.

For the specific case of inference in the dynamic Bradley-Terry model, ADF-type strategies offer an alternative to the mode-centred linearisation strategy of the Extended Kalman filter. This is potentially appealing, as such linearisation strategies are observed to be problematic for skewed likelihoods, which are commonplace in models with binary observations, as we have. In such settings, moment-matching approaches like ADF are observed to have improved robustness properties. See \cite{kuss2005assessing} for extensive discussion.

\section{Experimental Setup}

We here detail some global parameter choices for our experiments in Section 5. Experiments of the main text.

For the sigmoidal models, we run SMC with $J = 1000$ particles and the fHMM with $S = 500$ states. Naturally, increasing these resolution parameters can only increase accuracy at the cost of computational speed, whose complexities are laid out in Table 1. As mentioned previously, for the model \eqref{specific_SSM}, the parameters $\mu_0$ and $s$ are not identifiable given $\sigma_0, \tau$ and $\epsilon$; we therefore set $\mu_0 = 0$ and $s=1$. For the fHMM case, we note that the boundary conditions of the dynamics result in the scaling parameter $s_d$ being identifiable, and requires scaling with $S$. It would be possible to tune $s_d$ with parameter estimation techniques, but for ease of comparison with the continuous state-space approaches, we here fix it to $s_d = S/5$, which was found to work well in practice. This leaves the following parameters to be learnt from data: $K$ for Elo, $(\sigma_0, \tau, \epsilon)$ for Extended Kalman, TrueSkill2 and SMC, and $(\sigma_d, \tau_d, \epsilon_d)$ for fHMM. For the Extended Kalman approach, we use the state-space model \eqref{specific_SSM} with the logistic sigmoid function to match the Elo and Glicko approaches. TrueSkill2 requires the inverse probit sigmoid function which we also use for the SMC and fHMM approaches.

For the bivariate Poisson model with increased dimension, we again use SMC with $J=1000$ however are forced to decrease to $S=40$ states for the discrete model which has exponential scaling in the dimension of the single-player latent skill.

All code for replication can be found at \href{https://github.com/SamDuffield/abile}{github.com/SamDuffield/abile}.

\subsection{Data}

The data used for this article comprises of match results for professional tennis, football and chess, home and away goals scored for football and all accompanying match timestamps. Due to their broad utility, all data is available from a variety of public sources. For our simulations we use
\begin{itemize}
    \item \href{http://www.tennis-data.co.uk/alldata.php}{tennis-data.co.uk} for WTA tennis results and timestamps.
    \item \href{https://www.football-data.co.uk/englandm.php}{football-data.co.uk} for Premier League football results and timestamps.
    \item \href{https://github.com/martj42/international_results}{github.com/martj42/international\_results} for international football results and timestamps.
    \item \href{https://github.com/huffyhenry/forecasting-candidates}{github.com/huffyhenry/forecasting-candidates} for chess results and timestamps.
\end{itemize}

All of the data can be downloaded easily from the above sources, with helper functions found in \href{https://github.com/SamDuffield/abile/tree/main/datasets}{github.com/SamDuffield/abile/datasets}.

\end{document}